\documentclass[aps,showpacs,amsmath,amssymbd,dvips,showkeys,linenumbers,nofootbib,preprint]{revtex4}

\usepackage{graphicx}
\usepackage{color}
\usepackage[T1]{fontenc}

\DeclareMathOperator*{\diag}{\tt diag}

\def \be  {\begin{equation}}
\def \ee  {\end{equation}}
\def \ee  {\end{equation}}
\def \bea {\begin{eqnarray}}
\def \eea {\end{eqnarray}}

\begin{document}

\preprint{ECTP-2014-02\hspace*{0.5cm}and\hspace*{0.5cm}WLCAPP-2014-02}
\title{SU(3) Polyakov Linear Sigma-Model in an External Magnetic Field}

\author{Abdel~Nasser~Tawfik\footnote{http://atawfik.net/}}
\affiliation{Egyptian Center for Theoretical Physics (ECTP), MTI University, 11571 Cairo, Egypt}
\affiliation{World Laboratory for Cosmology And Particle Physics (WLCAPP), Cairo, Egypt}

\author{Niseem~Magdy} 
\affiliation{World Laboratory for Cosmology And Particle Physics (WLCAPP), 11571 Cairo, Egypt}
\affiliation{Brookhaven National Laboratory (BNL) - Department of Physics
P.O. Box 5000, Upton, NY 11973-5000, USA}

\begin{abstract}

In the present work, we analyse the effects of an external magnetic field on the chiral critical temperature $T_c$ of strongly interacting matter. In doing this, we can characterize  the magnetic properties of the quantum chromodynamics (QCD) strong interacting matter, the quark-gluon plasma  (QGP). We  investigate this in the framework of the SU(3) Polyakov linear sigma-model (PLSM). To this end, we implement two approaches representing two systems, in which the Polyakov-loop potential added to PLAMS either renormalized or non-normalized. The effects of Landau quantization on the strongly interacting matter is conjectures to reduce the electromagnetic interactions between quarks. In this case, the color interactions will be dominant and increasing, which - in turn - can be achieved by increasing of the Polyakov-loop fields.  Obviously, each of them equips us with a different understanding about the critical temperature under the effect of an external magnetic field. In both systems, we obtain a paramagnetic response. In one system, we find that $T_c$ increases with increasing the magnetic field. In the other one, $T_c$ significantly decreases with increasing the magnetic field.

\end{abstract}

\pacs{12.39.Fe, 12.38.Aw, 52.55.-s}
\keywords{Chiral Lagrangian, Quark confinement, Magnetic confinement and equilibrium}

\maketitle
\tableofcontents
\makeatletter
\let\toc@pre\relax
\let\toc@post\relax                 
\makeatother 

\section{Introduction}

It is believed that at high temperatures and densities there should be phase transition(s) between combined  nuclear matter and  the quark-gluon plasma (QGP), where quarks and gluons are no longer confined inside hadrons \cite{Rischke:2003mt}. The theoretical and experimental studies of QGP still represent a challenge to be faced by particle scientists. There are many heavy-ion experiments aiming to create that phase of matter and to study its properties like the Relativistic Heavy-Ion Collider (RHIC) at BNL, the Large Hadron Collider (LHC) at CERN, Nuclotron-based  Ion Collider facility (NICA) at JINR and the Facility for Antiproton and Ion Reaserch (FAIR) at GSI. From the theoretical point-of-view, there are - apart from Quantum Chromodynamic (QCD) and its numerical simulations - two main first-principle models, the Polyakov Nambu-Jona-Lasinio (PNJL) model  \cite{Fukushima:2003fw,Ratti:2005jh,Fukushima:2008wg} and a combination of the chiral linear-sigma model \cite{Gell Mann:1960} and the Polyakov loops (PLSM) or Polyakov quark meson model (PQM) for three quark flavors (two light plus one strange quarks) \cite{Schaefer:2008ax,Schaefer:2009ab,Mao:2010,Kahara:2008}. These models offer the theoretical framework to study some properties of QGP. Here, we are interested on the magnetic response of QGP.

With this regard, one the most significant researches nowadays is the one devoted to characterize  the QGP properties, for instance  the magnetic properties.  Many of these studies tackle the possible change in such properties when the strongly interacting system goes through phase transition(s) between hadronic and partonic phases and affected by an external magnetic field  \cite{Marco2010,Marco2011,Marco2011-2,Fraga2013,Jens13,Skokov2011}. To this end, the effect of the external magnetic field on the chiral condensates which reflect the behavior of the chiral phase-transition, is a promising approach \cite{lattice-delia}.  Also, the effect of the external magnetic field on the deconfinement order-parameter (Polyakov-loop) which reflects the behavior of the confinement-deconfinement phase-transition can be studied  \cite{lattice-delia}. To the external magnetic field, the strongly interacting system (hadronic or partonic) can response with magnetization, $M$, and  magnetic susceptibility, $\chi_M$  \cite{Claudio2013}. Both quantities characterize the magnetic properties of the system of interest. 

In an external magnetic field, the hadronic and partonic states are investigated in different models, such as the Hadron Resonance Gas (HRG) model \cite{G_Endrödi}, and other effective models \cite{Klevansky1989,Gusynin1995,Babansky1998,Klimenko1999,Semenoff1999,Goyal2000,Hiller2008,Ayala2006,Boomsma2010}. The NJL model \cite{Klevansky1992,Menezes2009,Menezes_2009}, the chiral perturbation theory \cite{Shushpanov1997,Agasian2000,Cohen2007}, the quark model  \cite{Kabat2002} and certain limits of QCD  \cite{Miransky2002} are also implemented. Furthermore, there are some studies devoted to the magnetic effects on the dynamical quark masses \cite{Klimenko2008}. The chiral magnetic-effect was studied in context of the PNJL model  \cite{Fukushima2010}. The lattice QCD calculations in an external magnetic field have been reported \cite{lattice-maxim,lattice-delia,Braguta:2011hq,Bali:2011qj,Bali:2012zg}. The PLSM was used to estimate the effects of the magnetic field on the strongly interaction matter \cite{Mizher:2010zb,Skokov2010,Marco2012}.

In the present work, we apply Landau theory (Landau quantization) \cite{Landau} in order to add restrictions to the quarks due to the existence of free charges in the plasma phase. We notice that the proposed configuration requires additional temperature to derive the system through the chiral phase-transition. Accordingly, we find that the value of the chiral condensates increase with increasing the external magnetic field. A few remarks are now in order. In many different calculations for the thermal behavior of the chiral condensates and the deconfinement order-parameter (Polyakov-Loop) using PNJL or NJL \cite{Marco2010,Marco2011,Marco2011-2}, the external magnetic field was not constant. Also, the relation between the critical temperatures of the chiral and confinement phase-transitions and the magnetic field was elaborated \cite{Gabriel2012}. Almost the same study was conducted in PLSM \cite{Fraga2013,Jens13,Skokov2011}. All these studies lead to almost the same pattern, the critical temperature of the chiral phase-transition increases with increasing the external magnetic field. But, the critical temperature of the confinement phase-transition behaves, oppositely. This latter behavior  agrees - to some extend - with the lattice QCD calculations \cite{lattice-delia}. Such an agreement was dominant till the lattice QCD calculations \cite{Bali:2011qj}, in which we find that the behavior of the chiral critical temperature with the magnetic field  is opposite, i.e., the chiral critical temperature decreases with increasing the magnetic field. In the present work, we study the effects of the external magnetic field on the phase transition using SU(3) PLSM. We observe that the results are almost the same as in the lattice QCD \cite{lattice-delia}. This result agrees well with many effective models. Also, we introduce explanation for the new lattice QCD calculations \cite{Bali:2011qj} and add some modifications to SU(3) PLSM in order to reproduce these calculations.    

In light of this, we recall that the PLSM is widely utilized to different frameworks and for different purposes. The LSM was introduced by Gell-Mann and Levy in 1960 \cite{Gell Mann:1960} long time before QCD was known to be the theory of strong interaction. Many studies have been performed on LSM like $\mathcal{O}(4)$ LSM \cite{Gell Mann:1960}, $\mathcal{O}(4)$  LSM at finite temperature \cite{Lenaghan:1999si, Petropoulos:1998gt} and $U(N_f)_r  \times U(N_f)_l$ LSM for $N_f=2$, $3$ or even $4$ quark flavors \cite{l, Hu:1974qb, Schechter:1975ju, Geddes:1979nd}. In order to obtain reliable results, Polyakov-loop corrections were added to LSM, in which information about the confining glue sector of the theory was included in form of Polyakov-loop potential which is to be extracted from pure Yang-Mills lattice simulations \cite{Polyakov:1978vu, Susskind:1979up, Svetitsky:1982gs,Svetitsky:1985ye}. So far, many studies were devoted to investigating the phase diagram and the thermodynamics of PLSM at different Polyakov-loop forms with two \cite{Wambach:2009ee,Kahara:2008} and three quark flavors \cite{Schaefer:2008ax,Mao:2010}. Also, the magnetic field effect on the QCD phase-transition and other system properties are investigated using PLSM \cite{Mizher:2010zb,Skokov2010,Marco2012}.

The present paper is organized as follows.  In section \ref{sec:approaches}, we introduce details about SU(3) PLSM under the effects of an external magnetic field. Section \ref{sec:Results} gives some features of the PLSM in an external magnetic field, such as the quark condensates, Polyakov loop, some thermal quantities and the magnetic phase-transition. Section \ref{sec:disc} is devoted to discussion and outlook. The conclusions are outlined in section \ref{sec:conclusion}.

\section{Approaches}
 \label{sec:approaches}

In this section, we introduce more details about the SU(3) PLSM under the effect of an external magnetic field. In the first part, we introduce detailed expression for SU(3) PLSM in an external magnetic field. The second part is devoted to the so called vacuum effect and renormalized Polyakov-loop.

\subsection{SU(3) PLSM  in an External Magnetic Field} 
\label{subsec:PLSM-in-MF}
The Lagrangian of LSM with $N_f =2+1$  quark flavors and $N_c =3$ color degrees of freedom, where the quarks couple to the Polyakov-loop dynamics,  was introduced in Ref. \cite{Schaefer:2008ax,Mao:2010},
\begin{eqnarray}
\mathcal{L}=\mathcal{L}_{chiral}-\mathbf{\mathcal{U}}(\phi ,\phi^* ,T), \label{plsm}
\end{eqnarray}
where the chiral part of the Lagrangian $\mathcal{L}_{chiral}=\mathcal{L}_q+\mathcal{L}_m$ is of $SU(3)_{L}\times SU(3)_{R}$ symmetry  \cite{Lenaghan,Schaefer:2008hk}. The Lagrangian with $N_f =2+1$ consists of two parts.  The first part stands for fermions, Eq. (\ref{lfermion}) with a flavor-blind Yukawa coupling $g$ of the quarks. The coupling between the effective gluon field and quarks, and between the magnetic field, $B$ and the quarks is implemented through the covariant derivative  \cite{Skokov2010}.
\begin{eqnarray}
\mathcal{L}_q &=& \sum_f \overline{\psi}_f(i\gamma^{\mu}
D_{\mu}-gT_a(\sigma_a+i \gamma_5 \pi_a))\psi_f, \label{lfermion} 
\end{eqnarray}
where the summation $\sum_f$ runs over the three flavors, $f=1, 2, 3$ for $u$-, $d$- and $s$-quark, respectively. The flavor-blind Yukawa coupling $g$ should couple the quarks to the mesons \cite{blind}. The coupling of the quarks to the Euclidean gauge field $A_{\mu}$  was discussed in Ref. \cite{Polyakov:1978vu,Susskind:1979up}. For the Abelian gauge field, the influence of the external magnetic field $A_{\mu}^{M}$ \cite{Mizher:2010zb} is given by the covariant derivative $ D_{\mu}$ \cite{Skokov2010},
\begin{eqnarray}
D_{\mu}&=& \partial_{\mu}-i A_{\mu}-i Q A_{\mu}^{M}, \label{covdiv} 
\end{eqnarray}
where $Q$ is a matrix defined by the quark electric charges $Q=\diag(q_u,q_d,q_s) $  for up,down and strange quarks, respectively.

The second part of chiral Lagrangian stands for the the mesonic contribution, Eq. (\ref{lmeson}), 
\begin{eqnarray}
\mathcal{L}_m &=&
\mathrm{Tr}(\partial_{\mu}\Phi^{\dag}\partial^{\mu}\Phi-m^2
\Phi^{\dag} \Phi)-\lambda_1 [\mathrm{Tr}(\Phi^{\dag} \Phi)]^2 \nonumber\\
&-& \lambda_2 \mathrm{Tr}(\Phi^{\dag}
\Phi)^2+c[\mathrm{Det}(\Phi)+\mathrm{Det}(\Phi^{\dag})]
+\mathrm{Tr}[H(\Phi+\Phi^{\dag})].  \label{lmeson}
\end{eqnarray}
In Eq. (\ref{lmeson}), $\Phi$ is a complex $3 \times 3$ matrix, which depends on the $\sigma_a$ and $\pi_a$ \cite{Schaefer:2008hk}, where $\gamma^{\mu}$ are the chiral spinors, $\sigma_a$ are the scalar mesons and $\pi_a$ are the pseudoscalar mesons. 
\begin{eqnarray}
\Phi= T_a \phi _{a} =T_a(\sigma_a+i\pi_a),\label{Phi}
\end{eqnarray}
where $T_a=\lambda_a/2$ with $a = 0, \cdots, 8$ are the nine generators of the U(3) symmetry group and $\lambda_a$ are the eight Gell-Mann matrices \cite{Gell Mann:1960}. The chiral symmetry is  explicitly broken  by $H$  
 \begin{eqnarray}
H = T_a h_a.\label{H}
\end{eqnarray}
$H$ is a $3\times 3$ matrix with nine parameters $h_a$. 

When taking into consideration that the spontaneous chiral symmetry breaking takes part in vacuum state, then a finite vacuum expectation value of the fields $\Phi$ and $\bar{\Phi}$ are conjectured to carry the quantum numbers of the vacuum \cite{Gasiorowicz:1969}. As a result, the diagonal components of the explicit symmetry breaking terms $h_0$, $h_3$  and $h_8$ should not vanish \cite{Gasiorowicz:1969}. This leads to exact three finite condensates $\bar{\sigma_0}$, $\bar{\sigma_3}$ and $\bar{\sigma_8}$ on one hand. On the other hand, $\bar{\sigma_3}$ breaks the isospin symmetry SU(2)  \cite{Gasiorowicz:1969}. To avoid this situation, we restrict ourselves to SU(3). This can be $N_f= 2+1$ \cite{Schaefer:2008hk} flavor symmetry breaking pattern. Correspondingly, two degenerate light (up and down) and one heavy quark flavor (strange) are assumed. Furthermore, the violation of the isospin symmetry is neglected. This facilitates the choice of $h_a$ ($h_0 \neq 0$, $h_3=0$ and $h_8 \neq 0$) for light and strange quarks.  Additional to these, five other parameters should be estimated. These are the squared tree level mass of the mesonic fields $m^2$, two possible quartic coupling constants $\lambda_1$  and $\lambda_2$, Yukawa coupling $g$ and a cubic coupling constant $c$. The latter models the axial $U(1)_A$ anomaly of the QCD vacuum. It is more convenient to convert the condensates $\sigma_0$ and $\sigma_8$ into a pure non-strange and strange parts  \cite{Mao:2010}. To this end, an orthogonal basis transformation from the original basis $\bar{\sigma_0}$ and $ \bar{\sigma_8}$ to the non-strange $\sigma_x$ and strange $\sigma_y$ quark flavor basis is required \cite{Kovacs:2006}. 
\bea
\label{sigms}
\left( {\begin{array}{c}
\sigma _x \\
\sigma _y
\end{array}}
\right)=\frac{1}{\sqrt{3}} 
\left({\begin{array}{cc}
\sqrt{2} & 1 \\
1 & -\sqrt{2}
\end{array}}\right) 
\left({ \begin{array}{c}
\sigma _0 \\
\sigma _8
\end{array}}
\right).
\eea

The second term in Eq. (\ref{plsm}), $\mathbf{\mathcal{U}}(\phi, \phi^*, T)$, represents the Polyakov-loop effective potential \cite{Polyakov:1978vu}, which is expressed by using the dynamics of the thermal expectation value of a color traced Wilson loop in the temporal direction  
\bea
\Phi (\vec{x})=\frac{1}{N_c} \langle \mathcal{P}(\vec{x})\rangle ,
\eea
Then, the Polyakov-loop potential and its conjugate read 
\begin{eqnarray}
\phi &=& (\mathrm{Tr}_c \,\mathcal{P})/N_c, \label{phais1}\\ 
\phi^* &=& (\mathrm{Tr}_c\,  \mathcal{P}^{\dag})/N_c, \label{phais2}
\end{eqnarray}
where $\mathcal{P}$ is the Polyakov loop.  This can be represented by a matrix in the color space \cite{Polyakov:1978vu} 
\begin{eqnarray}
 \mathcal{P}(\vec{x})=\mathcal{P}\mathrm{exp}\left[i\int_0^{\beta}d \tau A_4(\vec{x}, \tau)\right],\label{loop}
\end{eqnarray}
where $\beta=1/T$, is the inverse temperature and $A_4 = iA^0$ is called Polyakov gauge \cite{Polyakov:1978vu,Susskind:1979up}. The Polyakov loop matrix can be given  as a diagonal representation \cite{Fukushima:2003fw}. 

The coupling between the Polyakov loop and the quarks is given by the covariant derivative of $D_{\mu}=\partial_{\mu}-i A_{\mu}$  in PLSM Lagrangian, Eq. (\ref{plsm})  \cite{Mao:2010}.  It is apparent that the PLSM Lagrangian, Eq. (\ref{plsm}), is invariant under the chiral flavor group. This is similar to the original QCD Lagrangian \cite{Ratti:2005jh,Roessner:2007,Fukushima:2008wg}. In order to reproduce the thermodynamic behavior of the Polyakov loop for pure gauge case, we use a temperature-dependent potential $U(\phi, \phi^{*},T)$. This should agree with the lattice QCD simulations and have $Z(3)$ center symmetry as that of the pure gauge QCD Lagrangian \cite{Ratti:2005jh,Schaefer:2007d}. In case of vanishing chemical potential, then $\phi=\phi^{*}$ and the Polyakov loop is considered as an order parameter for the deconfinement phase-transition  \cite{Ratti:2005jh,Schaefer:2007d}. In the present work, we use $U(\phi, \phi^{*},T)$ as a polynomial expansion in $\phi$ and $\phi^{*}$ \cite{Ratti:2005jh,Roessner:2007,Schaefer:2007d,Fukushima:2008wg}
 \begin{eqnarray}
\frac{\mathbf{\mathcal{U}}(\phi, \phi^*, T)}{T^4}=-\frac{b_2(T)}{2} \phi ~ \phi^* -\frac{b_3
}{6}(\phi^3+\phi^{*3})+\frac{b_4}{4}(\phi ~ \phi^*)^2, \label{Uloop}
\end{eqnarray}
where 
\begin{eqnarray} \label{bb}
b_2(T)=a_0+a_1\left(\frac{T_0}{T}\right)+a_2\left(\frac{T_0}{T}\right)^2+a_3\left(\frac{T_0}{T}\right)^3. 
\end{eqnarray}
 
In order to reproduce the pure gauge QCD thermodynamics and the behavior of the Polyakov loop as a function of 
temperature, we use the parameters listed out in Tab. \ref{parameter}
\begin{table}[hbt]
\begin{center}

\begin{tabular}{c}
\hline  
$ a_0=6. 75 $,\qquad $ a_1=-1. 95 $, \qquad $ a_2=2. 625 $, \qquad
 $ a_3=-7. 44 $ \\  $ b_3 = 0.75 $ \qquad $b_4=7.5 $   \\ 
\hline 
\end{tabular} 
\caption{The potential parameters are adjusted to the pure gauge lattice QCD such that the equation of state and the Polyakov-loop expectation values are reproduced \cite{Ratti:2005jh}. \label{parameter}}
\end{center}
\end{table}


\subsection{Proposed modifications} 
\label{subsec:PLSM-in-MF-motivation}

We introduce some modifications based on various understandings and interpretations to the lattice QCD calculations \cite{Bali:2011qj}. Our interpretation for the Landau quantization effects on the strongly interacting matter is that they reduce the electromagnetic interactions between quarks. In this case, the color interactions will be dominant and increasing. Increasing color interactions can be achieved by an increase in the Polyakov-loop fields, which - in turn - makes the glounic  and Polyakov-loop potential dominant. We introduce two parts of modifications. The first one is related to the ultraviolet divergent vacuum contribution \cite{Schaffner2014}. The second one is relying on renormalization of the Polyakov-loop potential \cite{Falk2013,Fodor2006}. Accordingly, the  Lagrangian, Eq. (\ref{plsm}), reads
\begin{eqnarray}
\mathcal{L}=\mathcal{L}_{chiral}-\mathbf{\mathcal{U}}(\phi_R ,\phi_{R}^{*} ,T), \label{plsm-new}
\end{eqnarray}
where $ \mathcal{L}_{chiral} $ will be the same as we defined them in Eqs. (\ref{lfermion}) and (\ref{lmeson}). But here, we add a new definition for the Polyakov-loop potential based on Ref. \cite{Falk2013,Fodor2006}, 
\begin{eqnarray}
\frac{\mathbf{\mathcal{U}}(\phi_{R}, \phi_{R}^*, T)}{T^4}=-\frac{b_2(T)}{2} \phi_{R} ~ \phi_{R}^* -\frac{b_3
}{6}(\phi_{R}^3+\phi_{R}^{*3})+\frac{b_4}{4}(\phi_{R} ~ \phi_{R}^*)^2, \label{Uloop-new}
\end{eqnarray}
where $b_{2}(T)$ is defined in Eq. (\ref{bb}), and $\phi_{R}$ represents the renormalized Polyakov-loop field (not implemented in the calculations),
\begin{eqnarray}
\phi_R (T,B) &=& Z ~ \phi(T,B), \label{Pr-loop-new} 
\end{eqnarray}
where $Z$ is the renormalization factor,
\begin{eqnarray}
Z &=& \phi_0^{T_s/T} ~ \exp\left(\dfrac{U_{loop}\left(\phi_R ,\phi_{R}^{*} ,T_s\right)}{2T}\right)^{T_s/T},\\
\dfrac{U_{loop}(\phi_R ,\phi_{R}^{*} ,T_s)}{T^4} &=& - \frac{b_2(T_s)}{2}~\phi_{R}~ \phi_{R}^{*}-\frac{b_3
}{6}(\phi_{R}^3+\phi_{R}^{*3})+\frac{b_4}{4}(\phi_{R} ~ \phi_{R}^{*})^2,
 \label{Z-loop}
\end{eqnarray}
where $T_s$  and $\phi_0$ are free parameters assuring good agreements to the pure gauge lattice data. All parameters are listed in the Tab. \ref{par-new}.

\begin{table}[hbt] 
\begin{center}
\begin{tabular}{c}
\hline  
$ a_0=6. 75 $,\qquad $ a_1=-1. 95 $, \qquad $ a_2=2. 625 $, \qquad
 $ a_3=-7. 44 $ \\  $ b_3 = 0.75 $ \qquad $b_4=7.5 $ , \qquad
 $ T_0=187~MeV $ \\  $ T_s = 270~MeV $ \qquad $ \phi_0=0.99 $   \\ 
\hline 
\end{tabular} 
\caption{The potential parameters are adjusted to the pure gauge lattice data such that the equation of state and the Polyakov-loop expectation values are reproduced, except for $ T_s$ and $ \phi_0$ they are adjusted constants.\label{par-new}}
\end{center}
\end{table}

This expression for the Polyakov-loop, Eq. (\ref{Pr-loop-new}), has been chosen, because it has two free parameters, $T_s$ and $\phi_0$, which can be adjusted to get strong Polyakov-loop field $\phi$. Again this increases the glounic and Polyakov-loop potential and makes them dominant.

\subsection{Potential} 
\label{Potential}

By using the Lagrangians, Eqs. (\ref{plsm}) and (\ref{plsm-new}) in the mean filed approximation, Appendix \ref{appnd:1}, and under magnetic catalysis, Appendix \ref{appnd:2}, we can evaluate the two potentials as given in Eqs. (\ref{potential}) and (\ref{new-potential}), respectively.
\begin{eqnarray}
\Omega_{1}(T, \mu) &=& \frac{-T \mathrm{ln}
\mathcal{Z}}{V}=U(\sigma_x, \sigma_y)+\mathbf{\mathcal{U}}(\phi, \phi^*, T)+\Omega_{\bar{\psi}
\psi} (T;\phi,\phi^{*},B). \label{potential} \\
\Omega_{2}(T,B) &=& U(\sigma_x, \sigma_y)+\mathbf{\mathcal{U}}(\phi_{R},\phi_{R}^{*}, T)+\Omega_{\bar{\psi}
\psi} (T;\phi_{R},\phi_{R}^{*} ,B) + \Omega_{q \bar{q}}^{vac}(\sigma_x,\sigma_y). \label{new-potential}
\end{eqnarray}
The purely mesonic potential is given as
\begin{eqnarray}
U(\sigma_x, \sigma_y) &=& \frac{m^2}{2} (\sigma^2_x+\sigma^2_y)-h_x
\sigma_x-h_y \sigma_y-\frac{c}{2\sqrt{2}} \sigma^2_x \sigma_y \nonumber \\
&+& \frac{\lambda_1}{2} \sigma^2_x \sigma^2_y +\frac{1}{8} (2 \lambda_1
+\lambda_2)\sigma^4_x + \frac{1}{4} (\lambda_1+\lambda_2)\sigma^4_y. \label{Upotio}
\end{eqnarray}

For the potential, Eq. (\ref{potential}) and by using the mesonic potential, Eq. (\ref{Upotio}), the Polyakov-loop potential, Eq. (\ref{Uloop}) and Eq. (\ref{DR}), we determine the quarks and antiquarks contributions to the potential at a vanishing chemical potential but finite magnetic field
\begin{eqnarray} 
\Omega_{\bar{\psi} \psi}(T,B) &=& -2 \sum_{f} \dfrac{|q_{f}| B T}{2 \pi} \sum_{\nu=0}^{\infty} \int \dfrac{d p}{2 \pi} \left(2-1 \delta_{0n}\right)  \left\{ \ln \left[ 1+3\left(\phi+\phi^* e^{-\frac{E_f}{T}}\right)\times e^{-\frac{E_f}{T}}+e^{-3 \frac{E_f}{T}}\right] \right. \nonumber \\ 
&& \hspace*{30mm} \left.  +\ln \left[ 1+3\left(\phi^*+\phi e^{-\frac{E_f}{T}}\right)\times e^{-\frac{E_f}{T}}+e^{-3\frac{E_f}{T}}\right] \right\}.  \label{qqpotio}
\end{eqnarray}

For the potential, Eq. (\ref{new-potential}), and by using the mesonic potential, Eq. (\ref{Upotio})  and the Polyakov-loop potential, Eq. (\ref{Uloop-new}), the quark term contribution reads
\begin{eqnarray}
\Omega_{\bar{\psi} \psi}(T,B) &=& -2 \sum_{f} \dfrac{|q_{f}| B T}{2 \pi} \sum_{\nu=0}^{\infty} \int \dfrac{d p}{2 \pi} \left(2-1 \delta_{0n}\right)  \left\{ \ln \left[ 1+3\left(\phi_{R}+\phi_{R}^{*} e^{-\frac{E_f}{T}}\right)\times e^{-\frac{E_f}{T}}+e^{-3 \frac{E_f}{T}}\right] \right. \nonumber \\ 
&& \hspace*{30mm} \left.  +\ln \left[ 1+3\left(\phi_{R}^{*}+\phi_{R} e^{-\frac{E_f}{T}}\right)\times e^{-\frac{E_f}{T}}+e^{-3 \frac{E_f}{T}}\right] \right\}.  \label{new-qqpotio}
\end{eqnarray}
If the last term in Eq. (\ref{new-potential}) represents the fermionic vacuum loop contribution \cite{Schaffner2014}, then
\bea 
\Omega_{q \bar{q}}^{vac}(\sigma_x,\sigma_y) &=& -2 Nc \sum_{f} \int \dfrac{d^3 p}{(2 \pi)^3}  E_f 
= \dfrac{- Nc}{8 \pi^2}  \sum_{f} m_{f}^{4}\; \ln\left(\dfrac{m_f}{\Lambda_{QCD}}\right), \label{U-vacuum}
\eea 
where $N_c$ is the color degree of freedom and $\Lambda_{QCD}$ is the QCD energy scale.

\section{Results}
 \label{sec:Results}

In this section, the chiral condensates $\sigma_x$ and $\sigma_y$ and the confinement order-parameters $\phi$ and $\phi^*$ shall be extracted, Appendix \ref{appnd:1}. Then, we  study the thermal behavior of the condensates and order-parameters $\partial \sigma_f/\partial T$ and $\partial \phi/\partial T$ fluctuations under different external magnetic fields. These quantities shall be used to characterize the thermal dependence of further physical quantities. Finally, we map out the magnetic phase-diagram. In all these calculations,  Eqs. (\ref{potential}) and (\ref{new-potential}) shall be implemented.

\subsection{Phase transition: quark condensates and order parameters}
\label{subsec:condensates}

The thermal evolution of the chiral condensates, $\sigma_x$ and $\sigma_y$, and the order parameters, $\phi$ and $\phi^*$ as calculated from Eqs. (\ref{potential}) and (\ref{new-potential}) at vanishing chemical potential and finite magnetic field shall be estimated. Same calculation will be repeated for the derivatives of these parameters. Then, using the minimization conditions given in Eqs. (\ref{cond1}) and (\ref{cond2}), we can characterize the dependence of the potential on the three parameters, the temperature $T$, the magnetic field $B$ and the minimization parameter. This assures a minimum potential, as well. Apparently, these parameters depend on the temperature and the magnetic field. Additionally, we have other four parameters, $\sigma_x$, $\sigma_y$, $\phi $ and $\phi^*$ for each approach, Eqs. (\ref{potential}) and (\ref{new-potential}). Therefore, the minimization procedure should be repeated for each of these parameters, while the other parameters should remain fixed, i.e. {\it global minimum of other parameters}. Repeating this process, we get each parameter as a function of the temperature at different values of the magnetic field. 

\begin{figure}[htb]
\centering{
\includegraphics[width=5.cm,angle=-90]{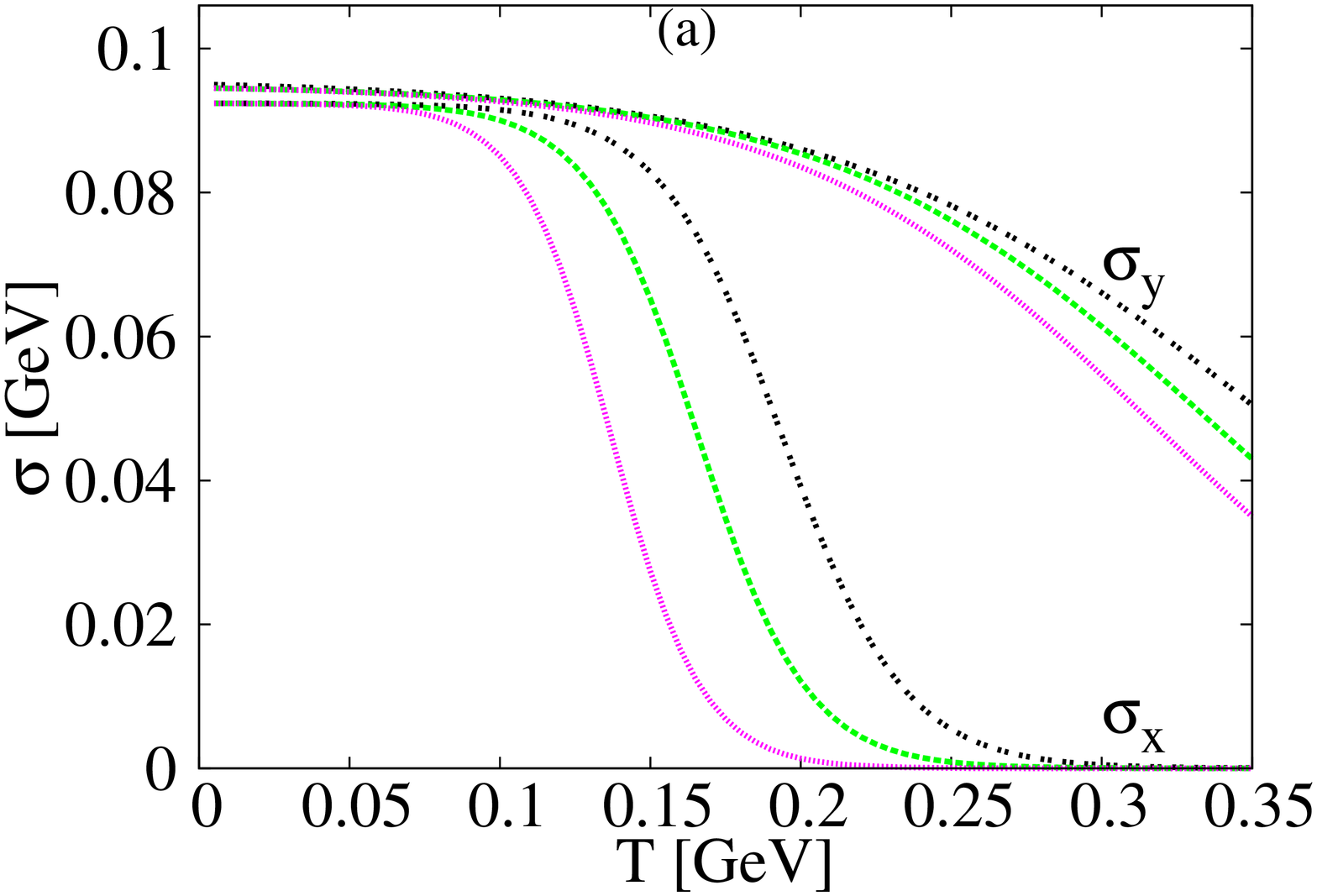}
\includegraphics[width=5.cm,angle=-90]{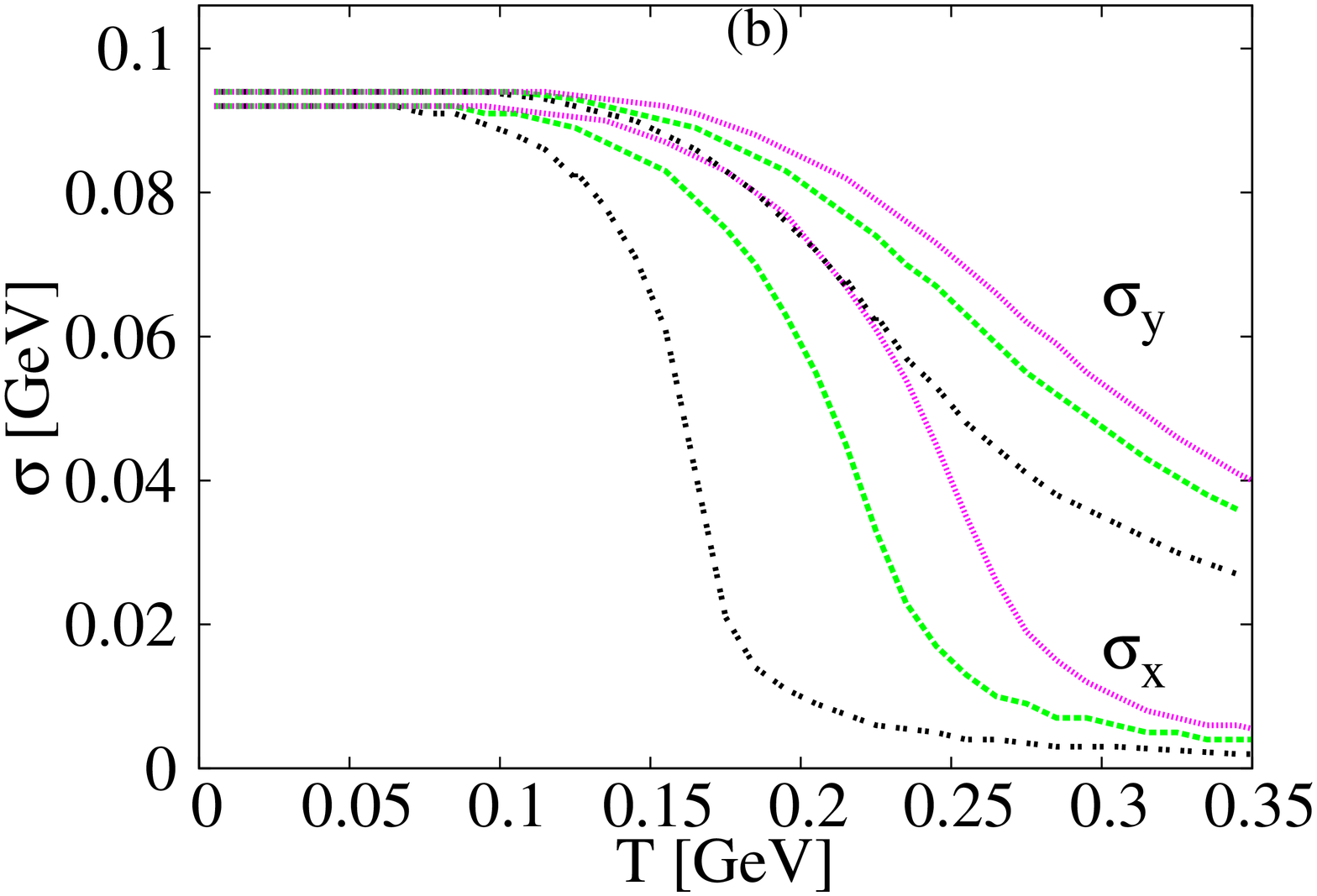}
\caption{(Color online) Left-hand panel (a): the strange and non-strange condensates of the renormalized approach, Eq. (\ref{potential}) at magnetic field $eB=0.019~$GeV$^2$ (double-dotted curve), $eB=0.2~$GeV$^2$ (dotted curve) and $eB=0.4~$GeV$^2$ (dashed curve). Right-hand panel (b): the same as in the left-hand panel (a) but for non renormalized approach, Eq. (\ref{new-potential}).   \label{fig:sig-with-t}}
}
\end{figure}

In the left-hand panel (a) of Fig. \ref{fig:sig-with-t}, the chiral condensate in the system controlled by Eq. (\ref{new-potential}) is given as a function of temperature at different magnetic fields. We notice here that increasing magnetic field $eB$ decreases the $\sigma_{x}$ and $\sigma_{y}$. The decrease of both quantities with the magnetic field is clear and gives a sign for the magnetic chiral phase-transition, Eq. (\ref{new-potential}). The behavior of the condensates seems to indicate that the chiral transition temperature decreases with the increase magnetic field.
In the right-hand panel (b) of Fig. \ref{fig:sig-with-t}, the chiral condensate calculated according to the potential, Eq. (\ref{potential}), is given as a function of temperature at different values of the magnetic field. It is clear that both strange and non-strange condensates increase as the magnetic field $eB$ increases. This behavior gives a clear signature for the magnetic chiral phase-transition for this approach, Eq. (\ref{potential}). That the chiral critical temperature increases with the magnetic field  agrees well with many studies using PNJL and PLSM  \cite{Marco2010,Marco2011,Marco2011-2}.

\begin{figure}[htb]
\centering{
\includegraphics[width=5.cm,angle=-90]{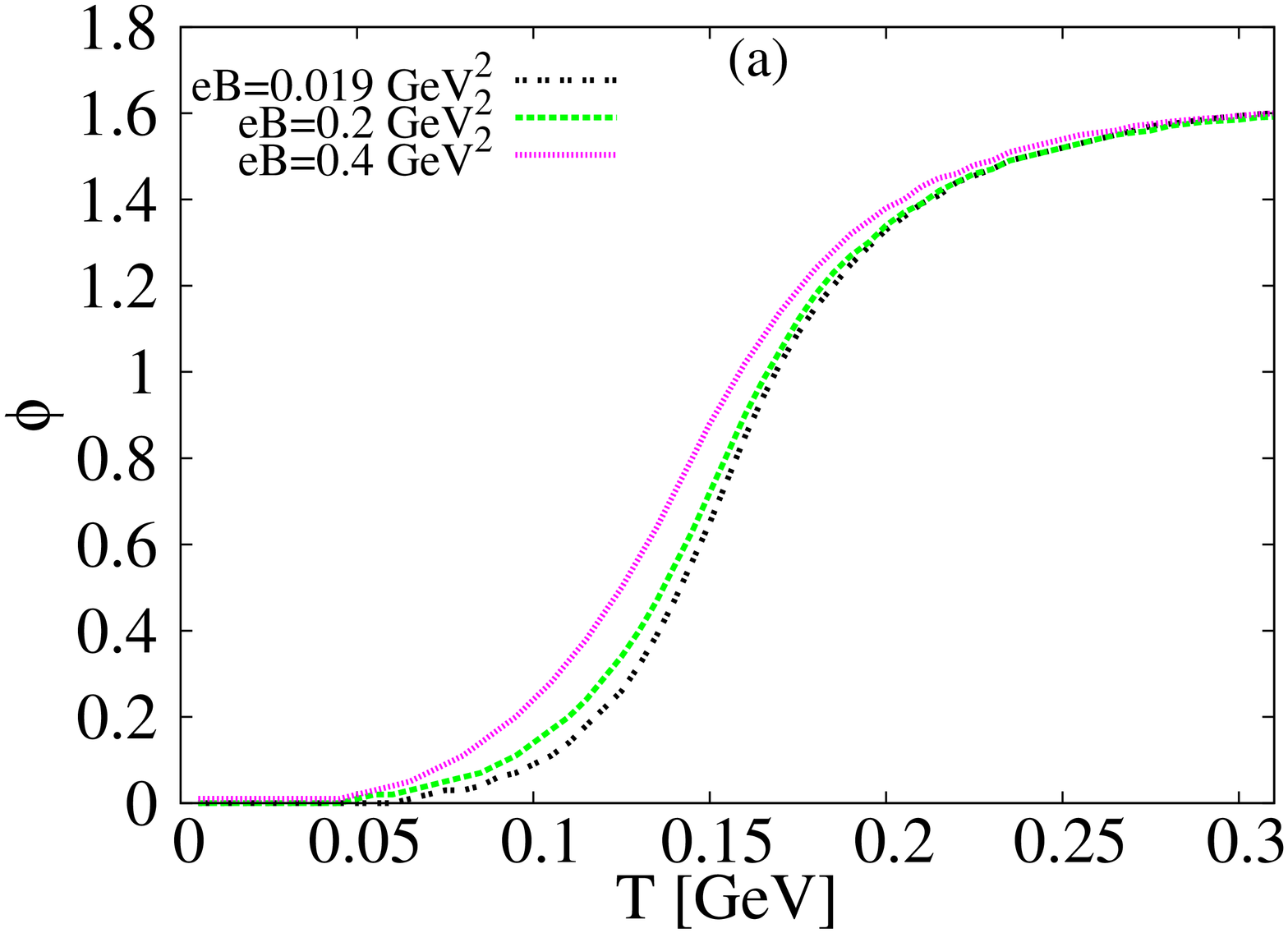}
\includegraphics[width=5.cm,angle=-90]{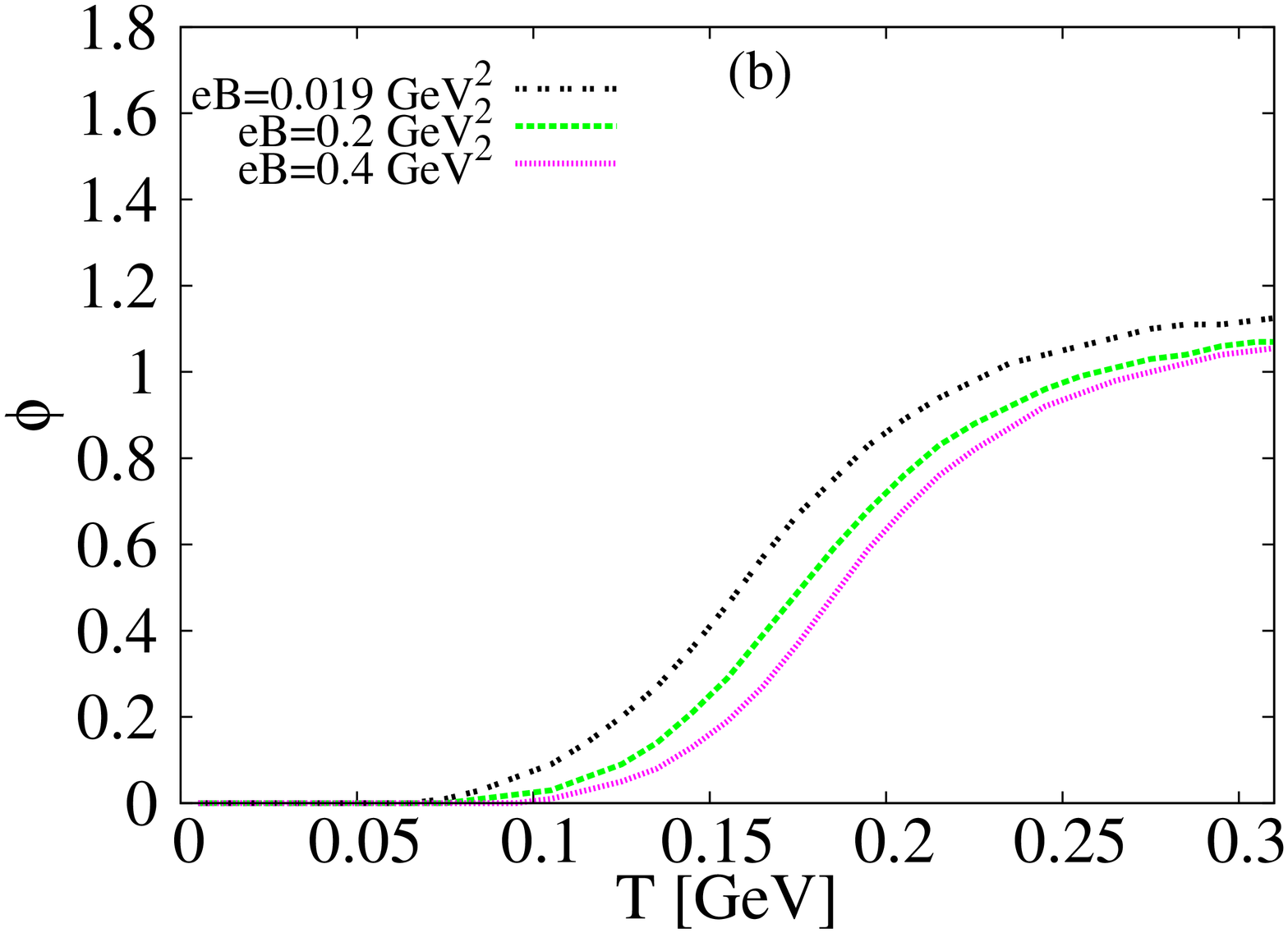}
\caption{(Color online) Left-hand panel (a): the Polyakov loop field $\phi$ of the renormalized approach, Eq. (\ref{new-potential}), at magnetic field $eB=0.019~$GeV$^2 $ (double-dotted curve), $eB=0.2~$GeV$^2$ (dotted curve) and $eB=0.4~$GeV$^2$ (dashed curve). Right-hand panel (b): the same as in the left-hand panel (a) but for non-renormalized approach, Eq. (\ref{potential}). \label{fig:fi-with-t}}
}
\end{figure}

In Fig. \ref{fig:fi-with-t}, the Polyakov loop field $\phi$ in the both systems Eqs. (\ref{potential}) and (\ref{new-potential}) is given as a function of temperatures at different values of the external magnetic field and at vanishing chemical potential $\phi=\phi^*$ and $\phi_{R}=\phi_{R}^{*} $. We notice that increasing $eB$ seems to affect the confinement phase-transition in different ways according to which approach is implemented.

In the left-hand panel (a) of Fig. \ref{fig:fi-with-t}, the Polyakov loop field $\phi$ in the system controlled by Eq. (\ref{new-potential}) is given as a function of temperatures at different values of the magnetic field. We find that increasing the magnetic field $eB$ increases the values of $\phi$, which is a clear signature for the confinement phase-transition in the PLSM potential, Eq. (\ref{plsm-new}).  Furthermore, the Polyakov loop field $\phi$ indicates that the deconfinement critical temperature increase with  increasing the magnetic field.  This behavior agrees with the recent lattice QCD calculations  \cite{Bali:2011qj}. 

In the right-hand panel (b) of Fig. \ref{fig:fi-with-t} draws the Polyakov loop field $\phi$ in the system controlled by  Eq. (\ref{potential}). We notice here that increasing $eB$ decreases the values of $\phi$. The decrease of this quantity give a signature for the confinement phase-transition in the PLSM potential, Eq. (\ref{potential}).  This behavior agrees with most previous studies in PLSM and PNJL under the effect of an external  magnetic field \cite{Marco2010,Marco2011,Marco2011-2}. The  Polyakov loop field $\phi$ indicates that the deconfinement critical temperature decreases with  increasing the magnetic field. This situation agrees with many studies using PNJL and PLSM  \cite{Marco2010,Marco2011,Marco2011-2}. 

\begin{figure}[htb]
\centering{
\includegraphics[width=5.cm,angle=-90]{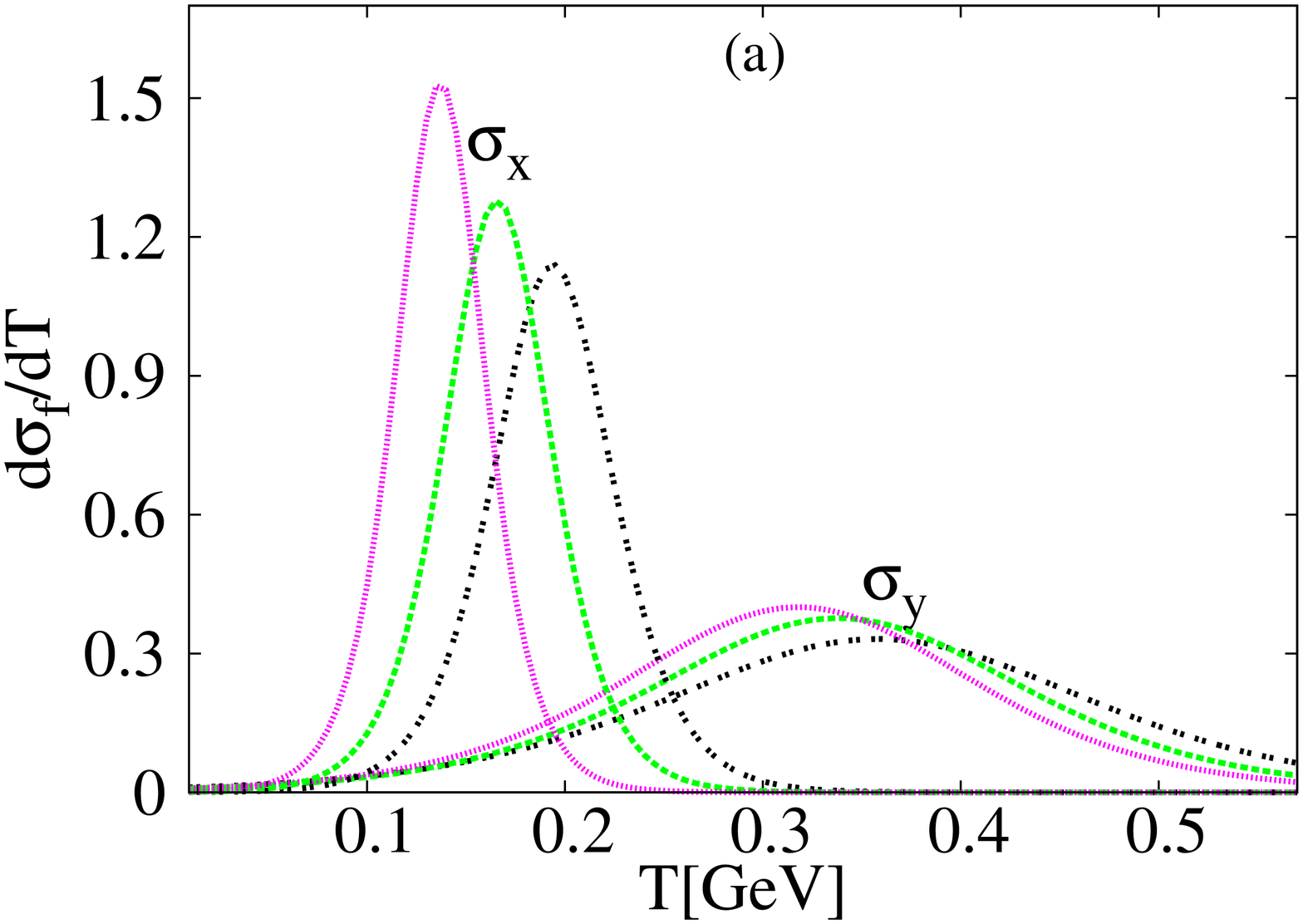}
\includegraphics[width=5.cm,angle=-90]{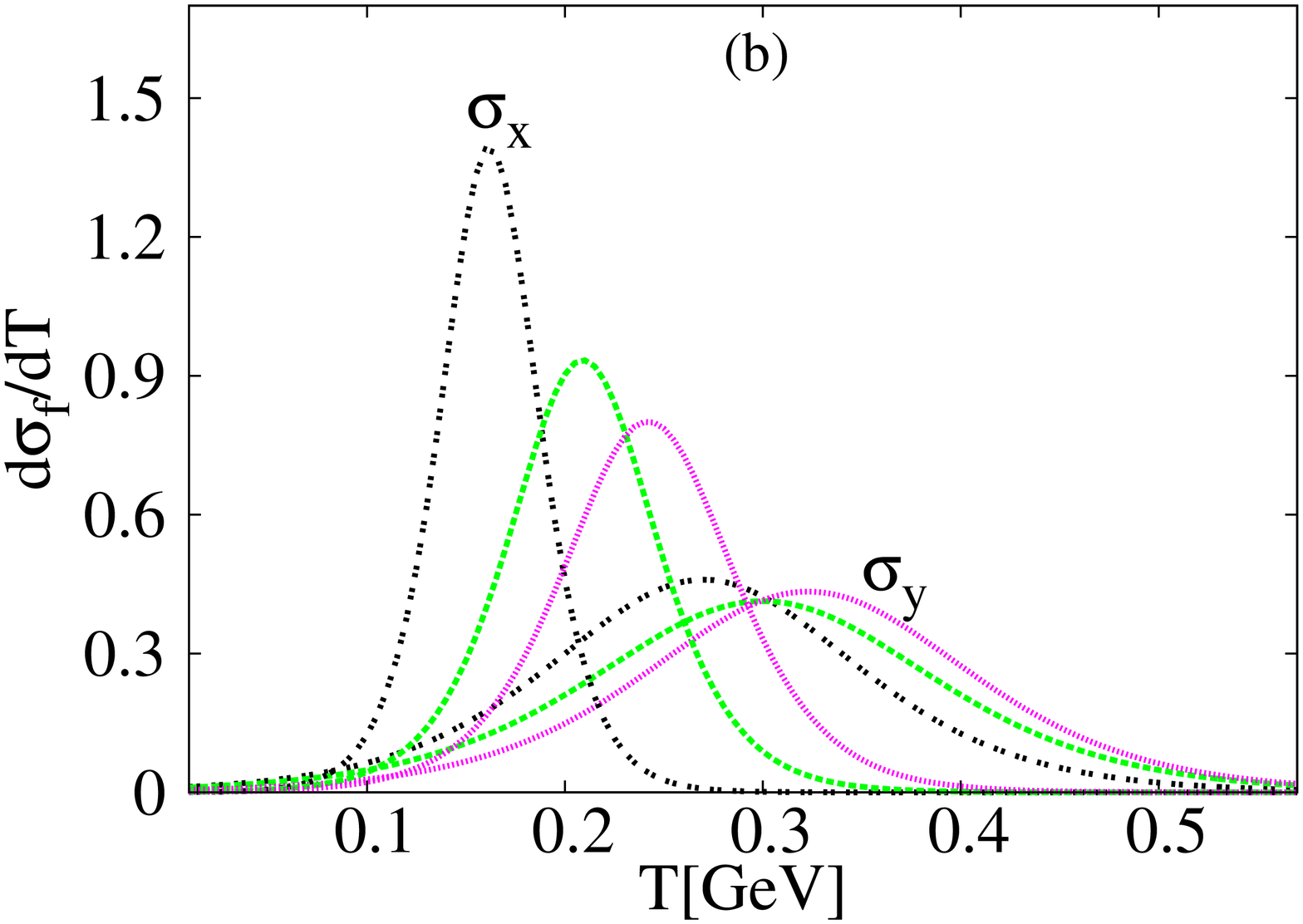}
\caption{(Color online) Left-hand panel (a): the derivative of strange and non-strange condensates $\partial \sigma_f/\partial T$ of the renormalized approach, Eq. (\ref{new-potential}) in the magnetic fields $eB=0.019~$GeV$^2 $ (double-dotted curve), $eB=0.2~$GeV$^2$ (dotted curve) and $eB=0.4~$GeV$^2$ (dashed curve). Right-hand panel (b): the same as in the left-hand panel (a) but for non-renormalized approach, Eq. (\ref{potential}.  \label{fig:Dsig-with-t}}
}
\end{figure}

In the left-hand panel (a) of Fig. \ref{fig:Dsig-with-t}, the thermal evolution of the derivative for the two chiral condensates for the approach given by Eq. (\ref{new-potential}) are given as functions of the temperatures at different values of the magnetic field. Again, we find that  increasing $eB$ decreases the  temperature, at which the peak, which represents the chiral phase-transition appears and simultaneously increases the height of the peak. The observed decrease in the temperature would be taken as a signature for the phase transition in the approach described by Eq. (\ref{new-potential}). Furthermore, the increase of the peak height likely reflects the effect of the external magnetic field on the quarks in this system. Such an effect adds some restrictions on the quarks. The most significant part here is the confinement term, $\phi$, Eq. (\ref{Pr-loop-new}). This term is elaborated in the left-hand panel (a) of Fig. \ref{fig:fi-with-t}. We find that it is larger than the counterpart term drawn in the right-hand panel (b) of Fig. (\ref{fig:fi-with-t}) of the system controlled by Eq. (\ref{potential}). As a consequence, we expect that the restrictions added to the system energy by Landau quantization through the external magnetic field should be induced by adding additional degree(s) of freedom to the gluons represented by Polyakov-loop potential. Obviously, this interpretation fits well with the lattice QCD results \cite{Bali:2011qj}. The decrease in both first derivatives for strange and non-strange chiral condensates with increasing  magnetic field is obvious and can be understood as a signature for phase transition in this approach.  

In the right-hand panel (b) of Fig. \ref{fig:Dsig-with-t}, the thermal evolution of the derivative of the two chiral condensates for the approach specified by Eq. (\ref{potential}) are given as functions of the temperatures at different values of the magnetic field. It apparent that increasing $eB$ increases the temperature, at which the peak takes place. This represents the chiral phase-transition in the approach given by Eq.(\ref{potential}). Furthermore, we notice that increasing $eB$ reduces the height of the peak, which likely reflect the effect of the external magnetic field on the quarks in the system controlled by Eq.(\ref{potential}). Through effect, one would understand that some restrictions on the quarks energy are likely added by the Landau quantization. For instance, the external magnetic field strengthens  this restrictions (reduces the peak). The results shown in the right-hand panel (b) of Figs. \ref{fig:Dsig-with-t} and \ref{fig:sig-with-t} agree with various studies using PLSM and PNJL, in which an external magnetic field was applied \cite{Marco2010,Marco2011,Marco2011-2,Fraga2013,Jens13,Skokov2011}.

\begin{figure}[htb]
\centering{
\includegraphics[width=5.cm,angle=-90]{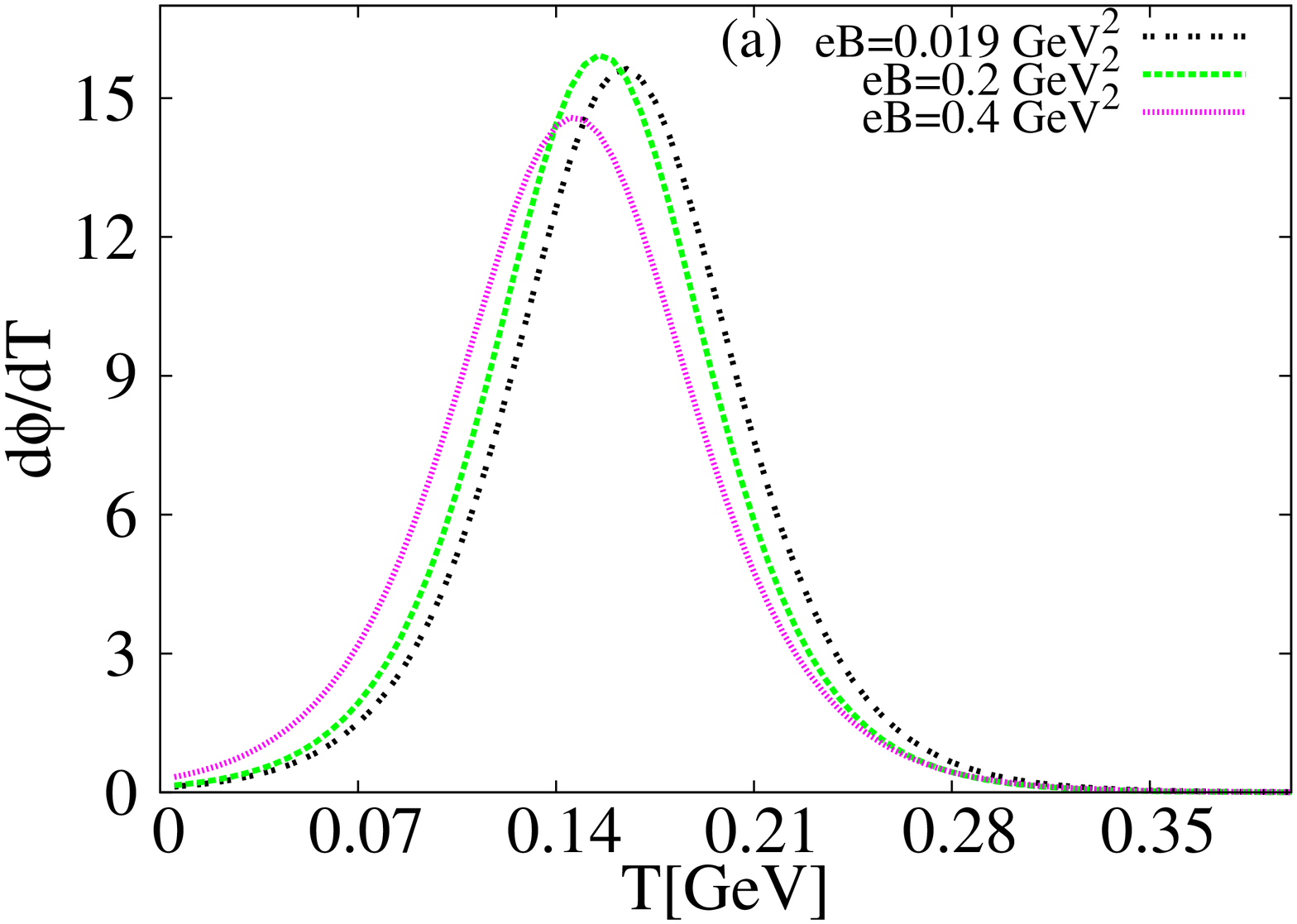}
\includegraphics[width=5.cm,angle=-90]{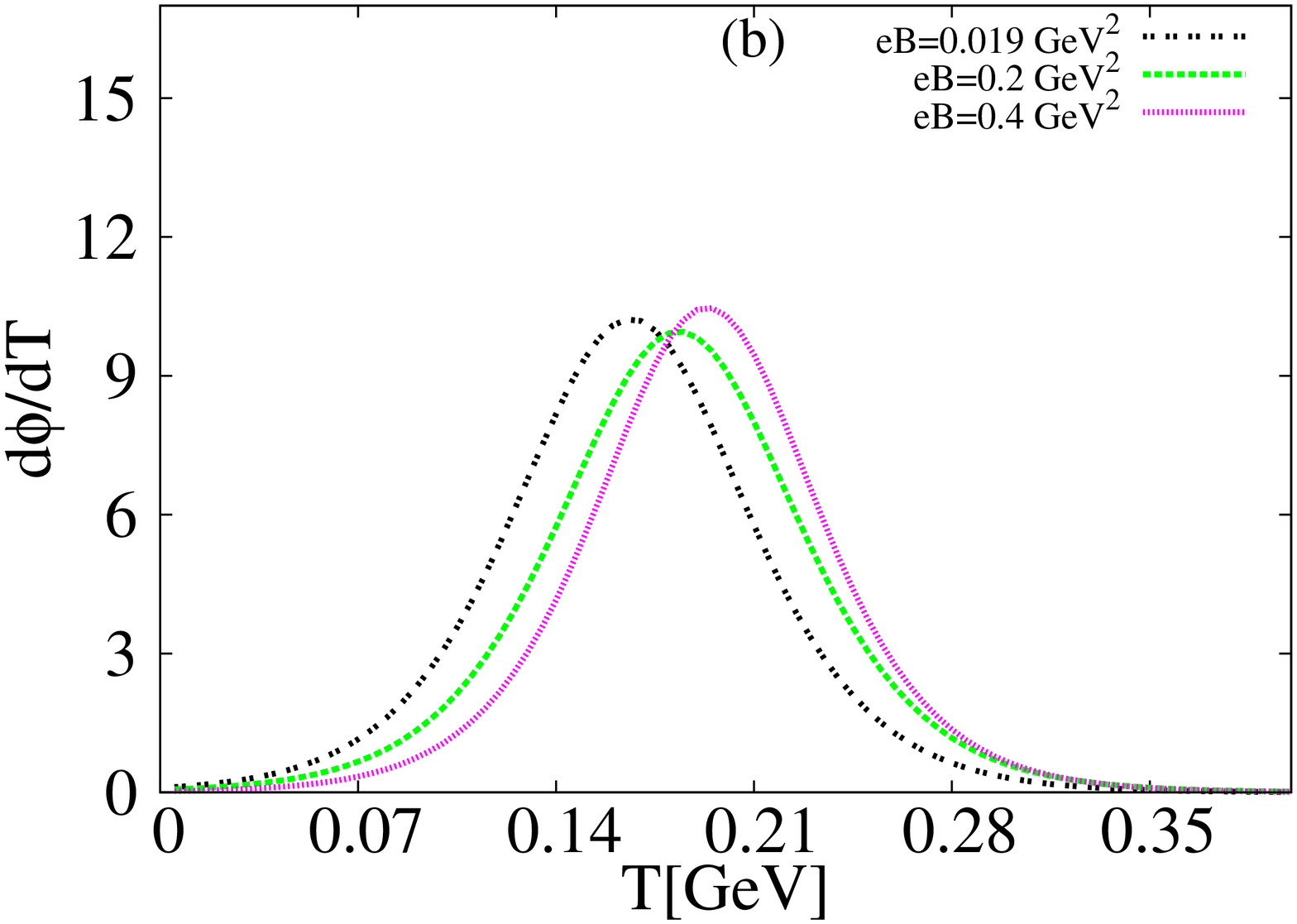}
\caption{(Color online) Left-hand panel (a): the Polyakov-loop potential $\phi$ of the renormalized approach, Eq. (\ref{new-potential}) at magnetic fields $eB=0.019~$GeV$^2$ (double-dotted curve), $eB=0.2~$GeV$^2$ (dotted curve) and $eB=0.4~$GeV$^2$ (dashed curve). Right-hand panel (b): the same as in the left-hand panel (a) but for non-renormalized approach, Eq. (\ref{potential}). \label{fig:Dfi-with-t}}
}
\end{figure}

In Fig. \ref{fig:Dfi-with-t}, the thermal evolution of the derivative for Polyakov-loop field $\phi$ with respect to $T$ in both systems controlled by Eqs. (\ref{potential}) and (\ref{new-potential})  is given as a function of $T$ at different values of the external magnetic field  and vanishing chimerical potential, i.e. $\phi =\phi^*$ even for the renormalized Polyakov loop potential. It is apparent that increasing $eB$ affects the confinement phase-transition in different ways.

In the left-hand panel (a) of Fig. \ref{fig:Dfi-with-t}, the thermal evolution of the derivative of $\phi$  in the system controlled by Eq. (\ref{new-potential}) is given as a function of  $T$ at different values of the external magnetic field. We find that Increasing $eB$ decreases the critical temperature, at which the peak representing the confinement phase-transition should takes place.
In the right-hand panel (b) of Fig. \ref{fig:Dfi-with-t}, the same as in the left-hand panel but for the system controlled by Eq. (\ref{potential}). We notice that increasing $eB$ increases the critical temperature.

\subsubsection{Subtracted chiral-condensate $\Delta_{ls}$}
\label{subsubsec:delta}

The subtracted chiral-condensate $\Delta_{ls}$ is a dimensionless quantity reflecting the difference between non-strange  and strange condensates 
\begin{eqnarray}
\Delta_{q,s}(T) &=& \dfrac{\langle\bar{q}q\rangle - \dfrac{m_q}{m_s} \langle\bar{s}s\rangle}{\langle\bar{q}q\rangle_{0} - \dfrac{m_q}{m_s} \langle\bar{s}s\rangle_{0}},\label{Eq:Delta}
\end{eqnarray}
where $\langle\bar{q}q\rangle$ ($\langle\bar{s}s\rangle$) being averaged non-strange (strange) condensate and $m_q$ ($m_s$) are non-strange (strange) mass. This quantity is apparently an order parameter for the chiral phase-transition. The latter is conjectured to take place, when $\Delta_{q,s}(T)$ rapidly decreases. 

\begin{figure}[htb]
\centering{
\includegraphics[width=5.cm,angle=-90]{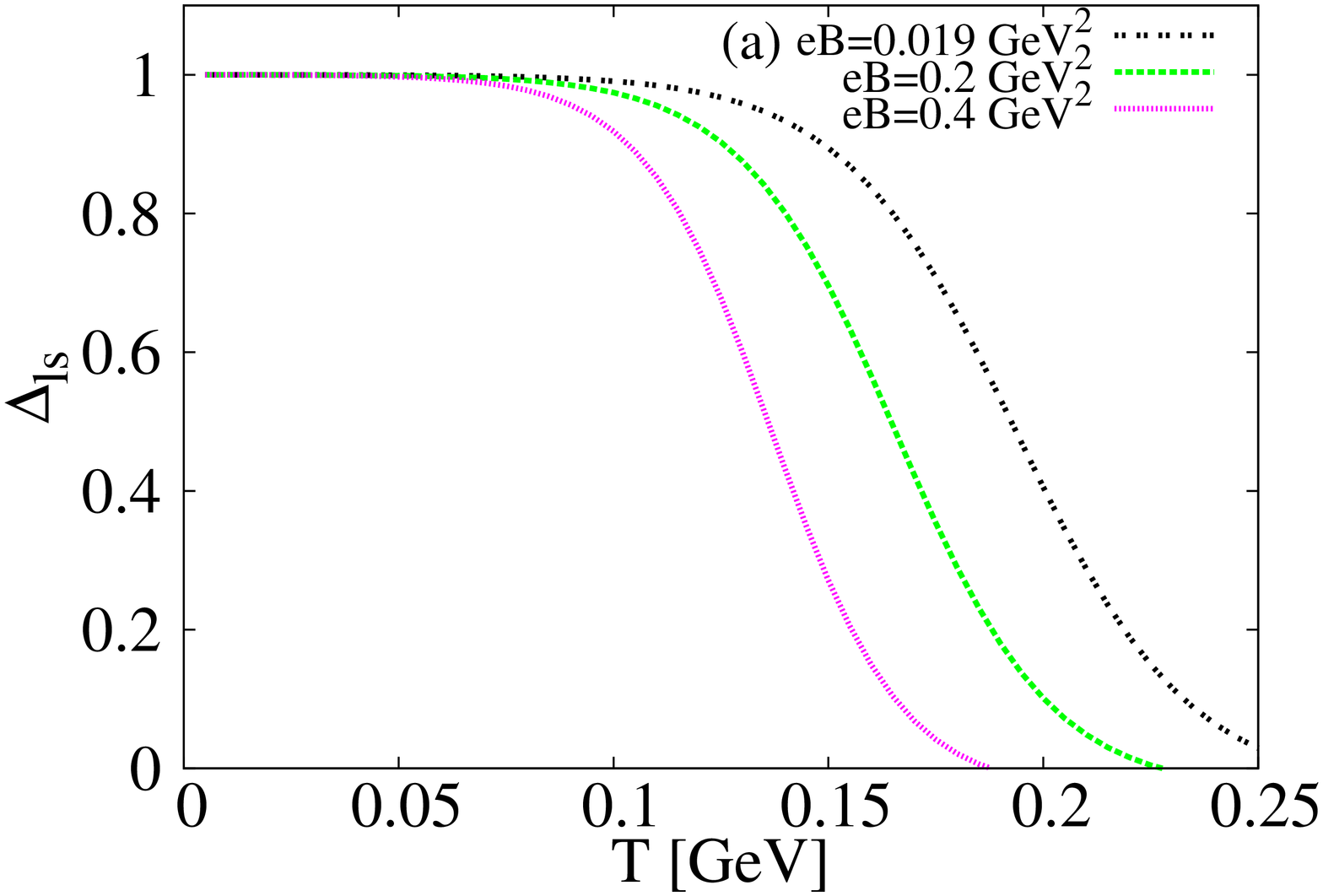}
\includegraphics[width=5.cm,angle=-90]{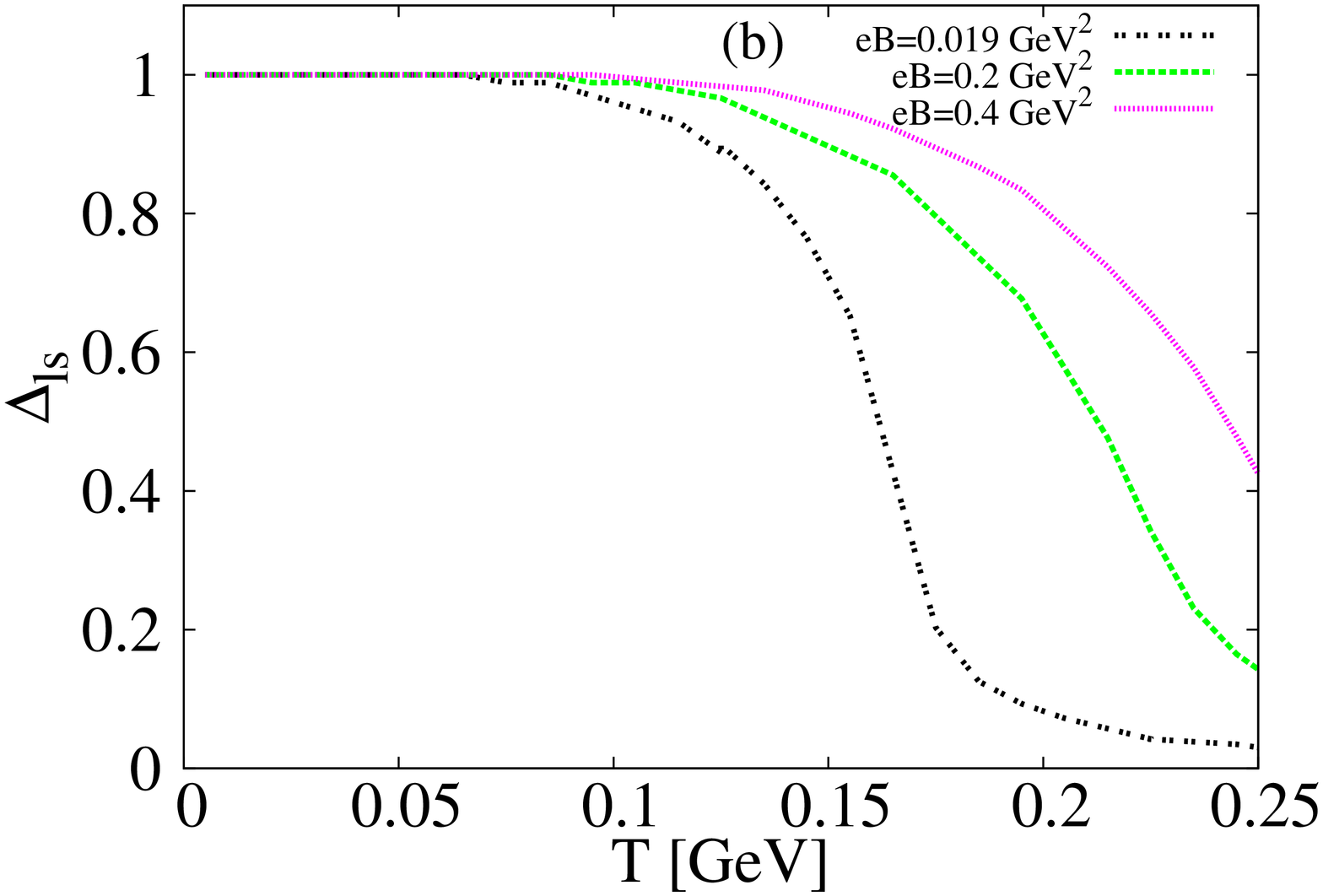}
\caption{(Color online) Left-hand panel (a): the subtracted chiral-condensate $\Delta_{ls}$ of the renormalized approach, Eq. (\ref{new-potential}), in external magnetic fields $eB=0.019~$GeV$^2$ (double-dotted curve), $eB=0.2~$GeV$^2$ (dotted curve) and $eB=0.4~$GeV$^2$ (dashed curve). Right-hand panel (b): the same as in the left-hand panel (a) but for non-renormalized approach, Eq. (\ref{potential}).  \label{fig:delta}}
}
\end{figure}

In Fig. \ref{fig:delta}, we draw the thermal evolution of $\Delta_{ls}$ at different values for the external magnetic field. In the left-hand panel (a), $\Delta_{ls}$ decreases as the external  magnetic field increases. Again, such a dependence was observed in the lattice QCD calculations \cite{Bali:2011qj}.  In right-hand panel (b), we find the $\Delta_{ls}$  increases as the external magnetic field increases. Almost the same dependence is present in other lattice QCD calculations \cite{lattice-delia}. So-far, we conclude that non-normalized and renormalized Polyakov-loop potential reproduces QCD-like (PNJL, LSM, etc.) and lattice QCD results, respectively. The renormalization of Polyakov-loop potential seems to play an essential role in interpreting the latest lattice QCD results. In section \ref{sec:lqcdCom}, we compare between two sets of lattice QCD calculations, \cite{lattice-delia} and \cite{Bali:2011qj}. With this regard, we also compare between non-normalized, Eq. (\ref{potential}), and renormalized, Eq. (\ref{new-potential}), approaches in section \ref{sec:PLFcom}.

\subsubsection{Comparison between lattice QCD calculations \cite{lattice-delia} and \cite{Bali:2011qj} }
\label{sec:lqcdCom}

A first-principle investigation for the properties of the deconfinement and chiral phase-transition in two-flavor QCD in the presence of a uniform background magnetic field using discretized pure gauge action and standard Wilson action was reported \cite{lattice-delia}. Different values of the bare quark mass, corresponding to pion masses in the range $200 - 480~$MeV, and magnetic fields up to $0.75~$GeV$^2$ were explored. It was concluded that the deconfinement and chiral critical-temperatures remain compatible with each other. Both raise very slightly as a function of the magnetic field.

In the lattice QCD calculations \cite{Bali:2011qj}, an improved gauge and smeared fermionic actions with $2+1$ flavors of quarks at the physical pion mass is implemented. The results are extrapolated to the continuum limit. Various values of the magnetic field, ranging from $0.1<\sqrt{eB}<1~$GeV are used. It was concluded that $T_c$ significantly decreases with increasing the magnetic field. This observation might conflict with \cite{lattice-delia} and various QCD-like calculations predicting an increasing $T_c$ with increasing magnetic field \cite{Marco2010,Marco2011,Marco2011-2,Fraga2013,Jens13,Skokov2011}. 

Should we want to differentiate between the two sets of lattice calculations, \cite{lattice-delia} and \cite{Bali:2011qj}, we can highlight the actions, the quark flavors and masses and the lattice sizes and spacings.

\subsubsection{Comparison between non-normalized, Eq. (\ref{potential}), and renormalized, Eq. (\ref{new-potential}), approaches}
\label{sec:PLFcom}

In the non-normalized approach, Eq. (\ref{potential}), the Polyakov-loop field is implemented in the LSM with $2+1$ quark flavors. As discussed, the magnetic effect is added through the covariant derivative. The quark potential is dominant just like that case without the magnetic filed \cite{TMD}. But there are some restrictions added to the quarks energy by Landau quantization through the magnetic effect.  This restrictions lower the value of the quark-potential term above $T_c$. This derives the system to require an additional amount of temperature in order to go through the hadron-quark phase transition, i.e. $T_c$ increases with increasing the external magnetic filed. This dependence agrees with the lattice QCD calculations \cite{lattice-delia}.

In the renormalized approach, Eq. (\ref{new-potential}), the restrictions added to the quarks  energy by the Landau quantization through the magnetic effect lower the quark-potential term above $T_c$. Furthermore, we add more degrees of freedom to the glounic term so that $T_c$ decreases with increasing magnetic field. This dependence - in tern - agrees with the lattice QCD calculations \cite{Bali:2011qj}.

The present work suggests two approaches qualitatively describing the two sets of lattice QCD calculations \cite{lattice-delia,Bali:2011qj}.

\subsection{Thermodynamic Quantities}
\label{subsec:thermo}
\begin{figure}[htb]
\includegraphics[width=6.cm,angle=-90]{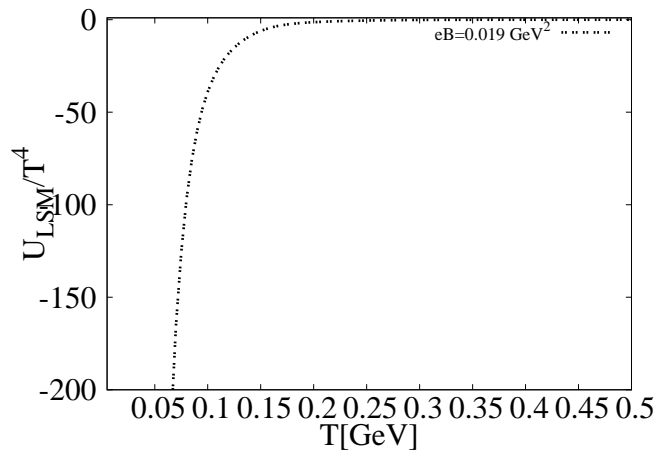}
\caption{(Color online) The thermal evolution of the mesonic potential of LSM is studied at vanishing chemical potential. Accordingly, this part of potential can be excluded, especially at high temperatures. 
\label{fig:LSMpotential} 
}
\end{figure}

In this section, we introduce some thermal quantities like pressure and trace anomaly. Estimating the contributions of the purely mesonic potential, Eq. (\ref{Upotio}), at various temperatures, Fig. \ref{fig:LSMpotential},  leads to the conclusion that it gets infinity at low temperature, but entirely vanishes at high temperatures. Therefore, this part of potential is only effective at very low temperatures. As the present study is performed at temperatures around the critical one, this part can be removed from the effective potentials given in Eqs. (\ref{potential}) and (\ref{new-potential}).  

\subsubsection{Pressure}
\label{subsubsec:presser}
The pressure density $P$ can obtained  from the grand canonical potentials, Eqs. (\ref{potential}) and (\ref{new-potential}), directly 
\begin{eqnarray}
P &=& - \Omega_1(T,B), \label{Pr1}\\
P &=& - \Omega_2(T,B).  \label{Pr2}
\end{eqnarray}
In the previous sections \ref{subsec:condensates}, we have estimated all parameters of the two fields. The two order parameters for two potentials are calculated, as well. Thus, we can now substitute all these into Eqs. (\ref{Pr1}) and (\ref{Pr2}).

\begin{figure}[htb]
\centering{
\includegraphics[width=5.cm,angle=-90]{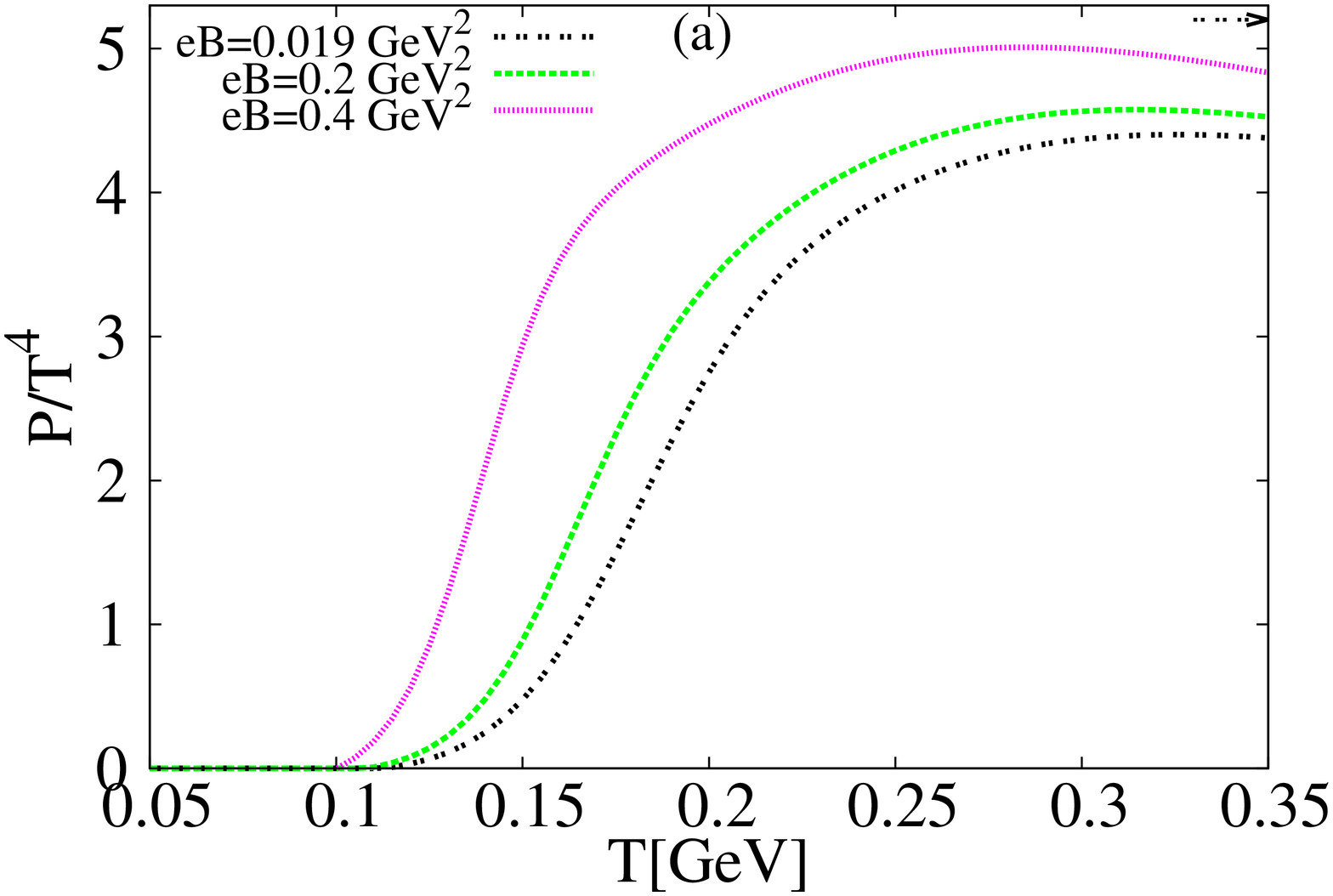}
\includegraphics[width=5.cm,angle=-90]{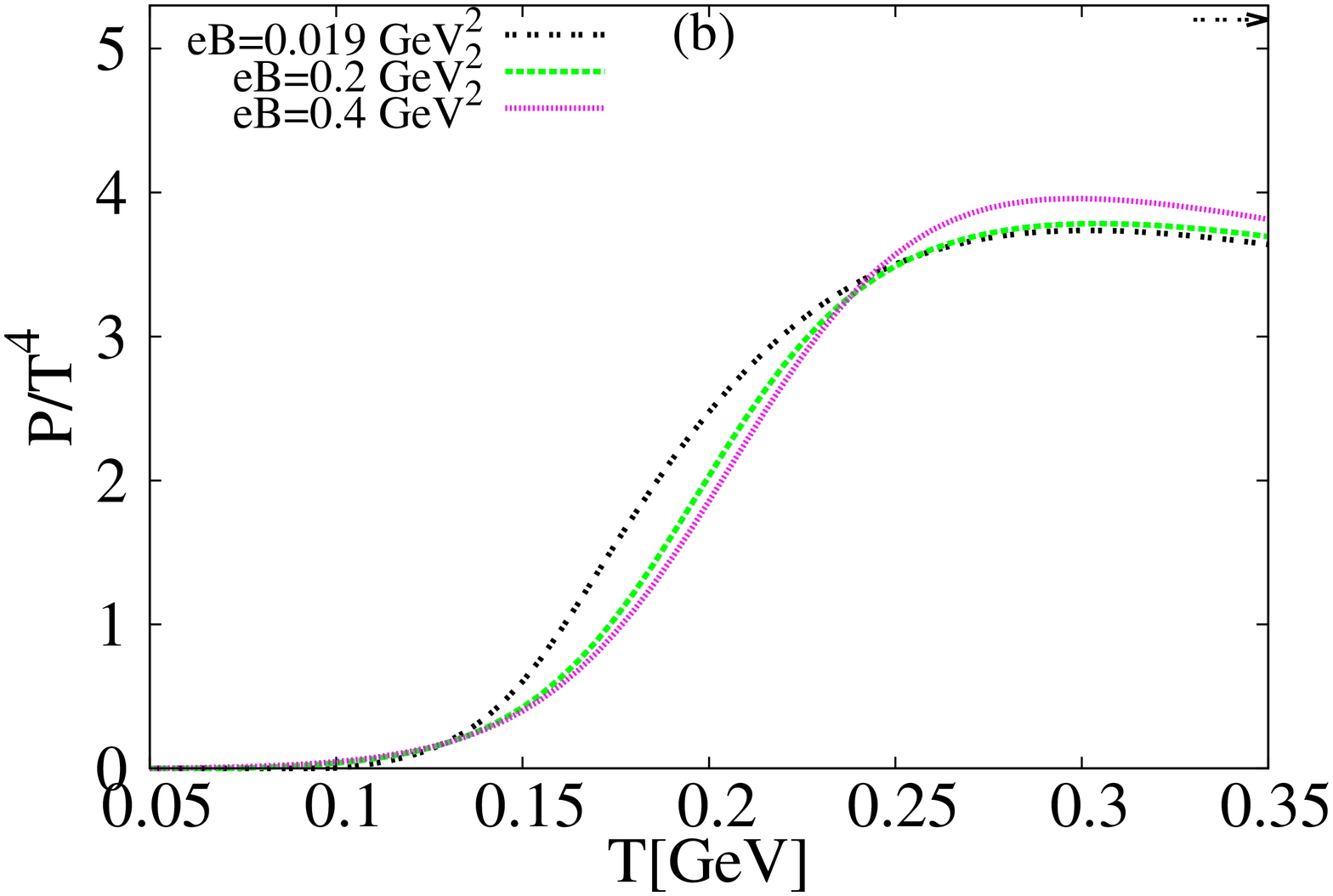}
\caption{(Color online) Left-hand panel (a): thermodynamic pressure of the renormalized approach, Eq. (\ref{new-potential}), at magnetic field $eB=0.019~$GeV$^2$ (double-dotted curve), $eB=0.2~$GeV$^2$ (dotted curve) and $eB=0.4~$GeV$^2$ (dashed curve). Right-hand panel (b): the same as in the left-hand panel (a) but for non-renormalized approach, Eq. (\ref{potential}). \label{fig:pr}}
}
\end{figure}

The thermal dependence of the pressure extracted from Eqs. (\ref{Pr2}) and (\ref{Pr1}) at different values for the external magnetic field, $eB=0.019~$GeV$^2$ (double-dotted curve), $eB=0.2~$GeV$^2$ (dotted curve) and $eB=0.4~$GeV$^2$ (dashed curve) is depicted in Fig. \ref{fig:pr}. It is obvious that the external magnetic field has a non-negligible effect on the pressure calculated according Eqs. (\ref{Pr2}) and (\ref{Pr1}). 

For instance for Eq. (\ref{Pr2}) in the left-hand panel (a) in Fig. \ref{fig:pr}, we notice that the critical temperature decreases as the external magnetic field increases. Also, the saturated region increases with increasing the field (below Stefan-Boltzmann limit). This behavior is close to the results of the PLSM in absence of an external magnetic field  \cite{TMD,Mao:2010}. The saturated region is similar to that from the PLSM without external magnetic field. This can be explained by the restrictions added to the quarks energy  through the magnetic field. Furthermore, we add more degrees of freedom to the glouns. This likely increases the pressure.

In the right-hand panel (b) in Fig. \ref{fig:pr}, the approach given in Eq. (\ref{Pr1}), the critical temperature increases as the magnetic field increases.  Also, the saturated part at large temperature gets higher with increasing the magnetic field. Nevertheless, it remains below the Stefan-Boltzmann limit. It remarkable that the same behavior is observed in PLSM without external magnetic field  \cite{TMD,Mao:2010}. The decrease in the height of the saturated region might be explained by the restrictions added to the quarks energy. The restriction effects become clear in the right-hand panel (b) of Fig. \ref{fig:pr},  Eq. (\ref{potential}). 

The potential in both Eqs. (\ref{potential}) and (\ref{new-potential}) is inverse-exponentially dependent on $T$ and $eB$. The quantity $2 \nu |q_f| eB $ increases with increasing $T$. At very large $T$, the dependence of exponential term on $T$ becomes very small. 

\subsubsection{Trace anomaly}
\label{subsubsec:trace}
The trace anomaly of the energy-momentum tensor $\mathcal{T}^{\nu \theta}$ also known as interaction measure reads
\begin{eqnarray} 
\Delta_{1} &=&  \dfrac{ \mathcal{T}_{1}^{\nu \theta} }{T^4} = 
T \frac{\partial}{\partial T} \; \frac{P_{1}}{T^4}, \label{tr1} \\
\Delta_{2} &=& \dfrac{ \mathcal{T}_{2}^{\nu \theta} }{T^4} = 
T \frac{\partial}{\partial T} \; \frac{P_{2}}{T^4}. \label{tr2}
\end{eqnarray}
Together with the pressure, we can now estimate almost all thermodynamic quantities.

\begin{figure}[htb]
\centering{
\includegraphics[width=5.cm,angle=-90]{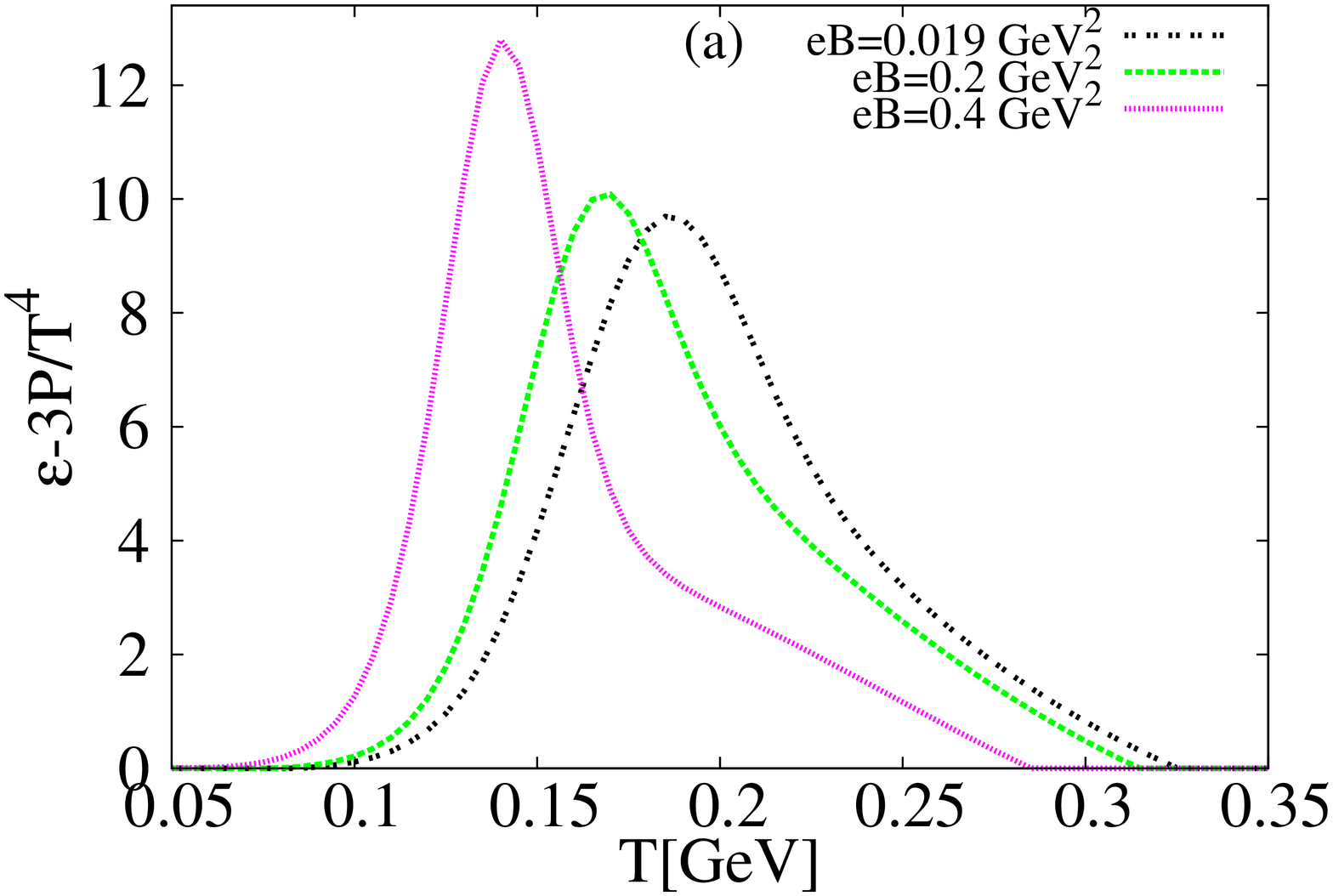}
\includegraphics[width=5.cm,angle=-90]{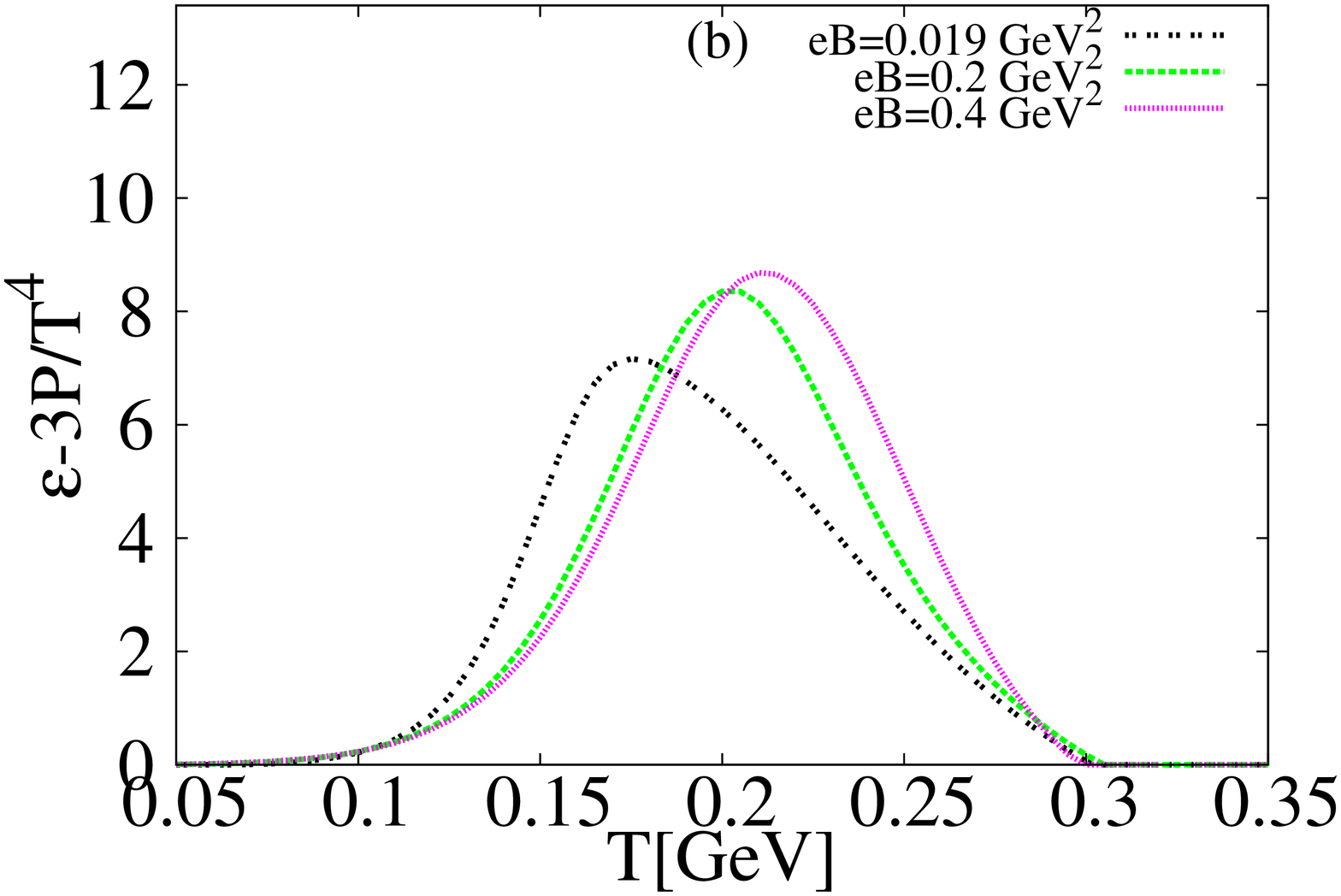}
\caption{(Color online) Left-hand panel (a): thermodynamic trace-anomaly  of the renormalized approach, Eq. (\ref{new-potential}), at various values of the external magnetic field $eB=0.019~$GeV$^2$ (double-dotted curve), $eB=0.2~$GeV$^2$ (dotted curve) and $eB=0.4~$GeV$^2$ (dashed curve). Right-hand panel (b): the same as in the left-hand panel (a) but for non-renormalized approach, Eq. (\ref{potential}). \label{fig:tr}}
}
\end{figure}

For the approach given by Eq. (\ref{tr2}), we notice that the critical temperature decreases with increasing the external magnetic field, left-hand panel (a) of Fig. \ref{fig:tr}. The opposite is observed for the other approach, Eq. (\ref{tr1}) in the right-hand panel (b).  It is worthwhile to mention that the phase transition is defined at the peak of the trace anomaly for both approaches, Eqs. (\ref{tr1}) and (\ref{tr2}).

\subsection{Magnetic susceptibility  $\chi_{m}$}
\label{subsubsec:susceptibility}
The response of the free energy density $f$ to an external magnetic  field represents another fundamental property of the strongly interacting QCD matter 
\begin{eqnarray} \label{f1}
 f &=& -\dfrac{T}{V} \ln\left(\mathcal{Z}\right),
\end{eqnarray}
where $\mathcal{Z}$ is the partition function of the system and $V$ the three-dimensional volume. The magnetic susceptibility of the strongly interacting QCD matter is given by the second derivative for $f$ with respect to the magnetic field, $eB$
\begin{eqnarray} \label{ch1}
\chi_{m} &=& \left. -\dfrac{\partial^{2} f}{\partial (eB)^2}\right|_{eB=0},
\end{eqnarray}
which is a dimensionless quantity. Here $e>0$ denotes the elementary charge. A positive susceptibility indicates a decrease in the free energy $f$  due to the external magnetic field. This is know as a paramagnetic response. On the other hand, negative $\chi_{m}$  is referred to diamagnetism. Clearly, the sign of $\chi_{m}$ is a fundamental property of the strongly interacting QCD matter.

\begin{figure}[htb]
\centering{
\includegraphics[width=5.cm,angle=-90]{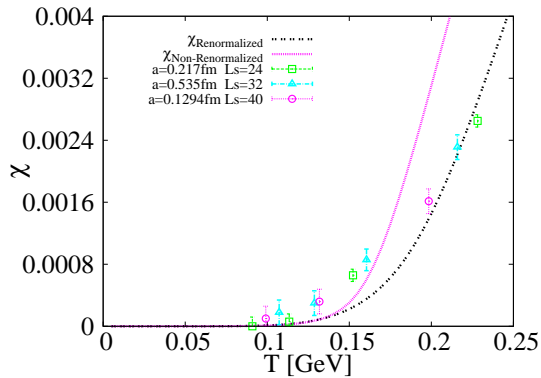}
\caption{(Color online) The magnetic susceptibility $\chi$ of the renormalized approach, Eq. (\ref{new-potential}), (dotted curve) and for non-renormalized approach, Eq. (\ref{potential}), (dashed curve). The symbols represent the lattice QCD results at different lattice spacing $a$ and sizes $L_s$ \cite{Claudio2013}. \label{fig:chim}}
}
\end{figure}

In Fig. \ref{fig:chim}, we find that both susceptibilities (from the two potentials Eqs. (\ref{potential}) and (\ref{new-potential})) represent a paramagnetic system, i.e. $\chi_{m}>0$. The susceptibility calculated from the renormalized potential, Eq. (\ref{new-potential}) agrees fairly with lattice QCD calculations \cite{Claudio2013}, especially at high $T$, while the susceptibility due to the non-renormalized potential, Eq. (\ref{potential}) (dashed line) fits well the lattice QCD calculations  \cite{Claudio2013}, especially at low $T$. These lattice QCD simulations implement different lattice spacings, $a$ and sizes $L_s$. The agreement with the lattice calculations apparently highlights correctness of the proposed approach, which seems to reproduce the first-principle lattice QCD calculation, especially the magnetic susceptibility calculated from the renormalized potential, Eq. (\ref{new-potential}). Furthermore, we find that the latter is closer to the lattice calculations \cite{Claudio2013} than the one calculated from the non-normalized potential, Eq. (\ref{potential}).

\subsection{Magnetic chiral phase-transition}
\label{subsec:phase-T}
\begin{figure}[htb]
\centering{
\includegraphics[width=5.cm,angle=-90]{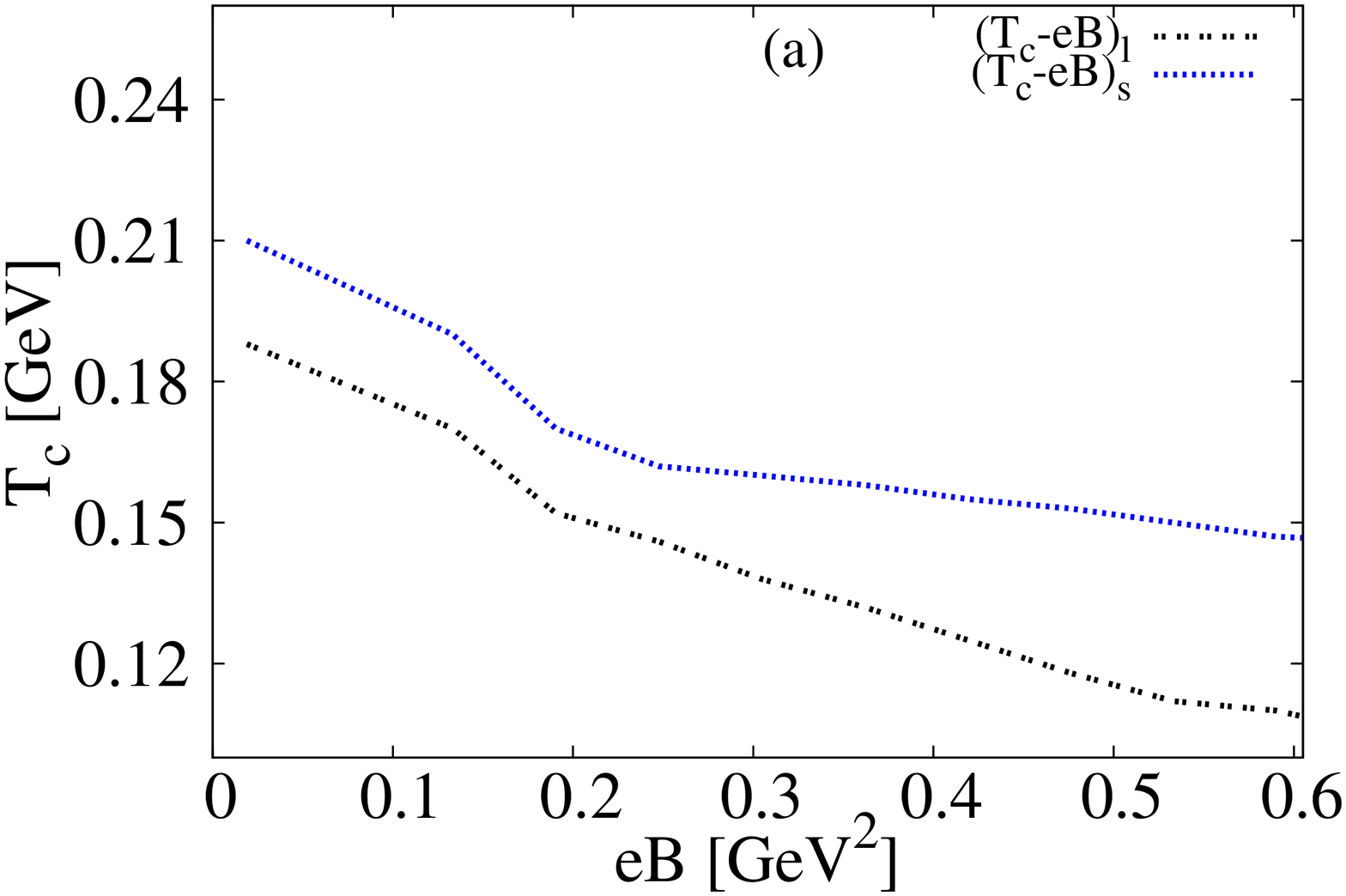}
\includegraphics[width=5.cm,angle=-90]{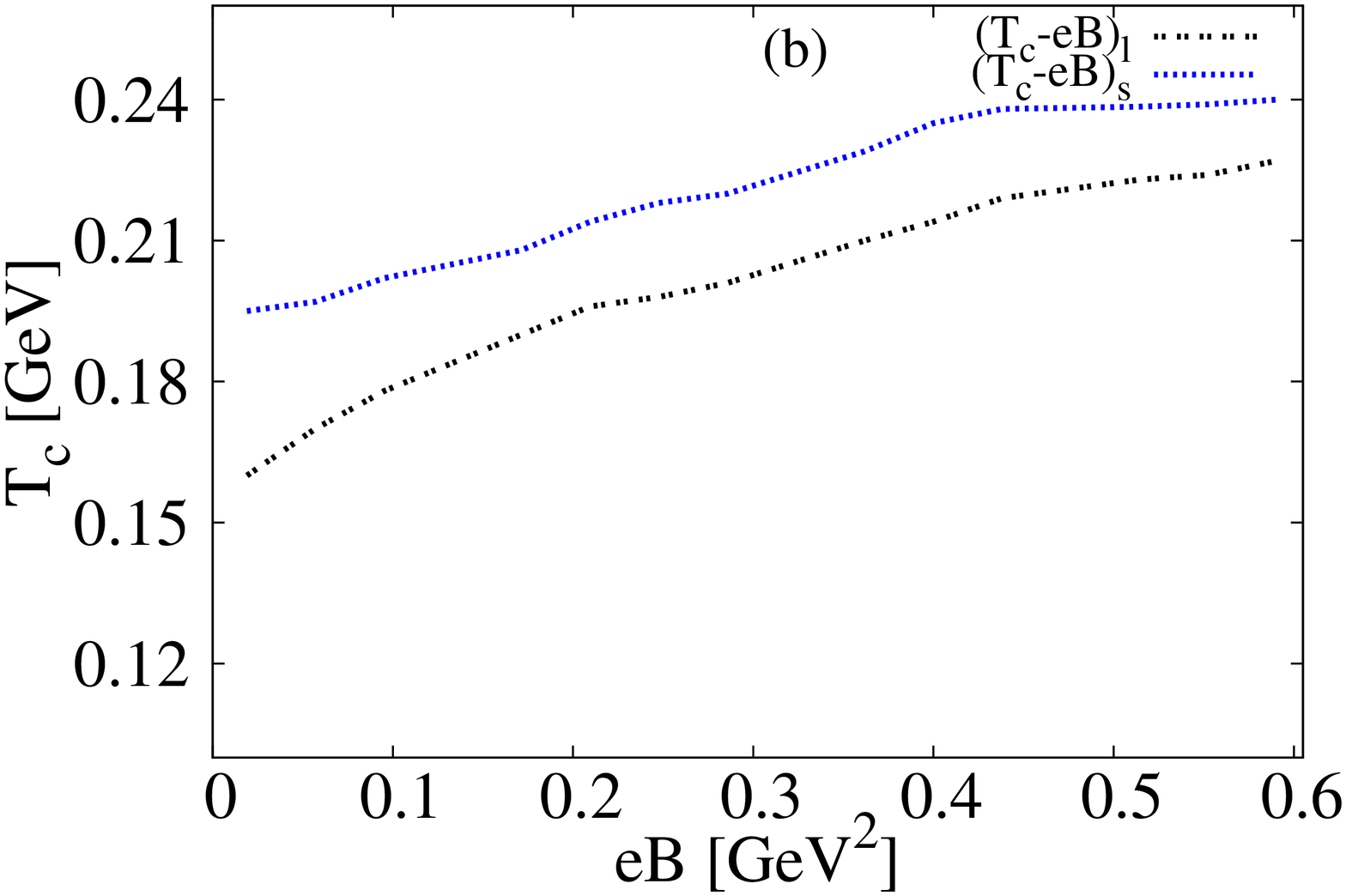}
\caption{(Color online) Left-hand panel (a): the magnetic chiral phase-diagram in an external magnetic field. The critical temperature $T_c$ is given in dependence on $eB$ from the renormalized approach, Eq. (\ref{new-potential}), for light chiral condensate (double-dotted curve) and for strange chiral condensate (dotted curve). Right-hand panel (b): the same as in the left-hand panel (a) but for the non-renormalized approach, Eq. (\ref{potential}). \label{fig:TP1}}
}
\end{figure}

In this section we introduce the magnetic chiral phase-transition under the effect of an external magnetic field for light and strange quarks. There are various procedures to estimate the chiral phase-transition \cite{TMD}. For instance, the peak of the first derivative of the condensates as shown in Fig. (\ref{fig:Dsig-with-t}) and the intersection between the condensates and the Polyakov loop. Since we have two potentials, Eqs. (\ref{potential}) and (\ref{new-potential}), we analyze each of them, separately, in Fig. (\ref{fig:TP1}).  

In left-hand panel (a) of  Fig. \ref{fig:TP1}, the magnetic chiral phase-diagram, $T$ vs. $eB$, for strange and light quarks from the renormalized approach described by Eq. (\ref{new-potential}) is presented. It is clear that $T_c$ decreases with $eB$ for both strange and light quarks. Qualitatively, this results agree well with the lattice QCD calculations \cite{Bali:2011qj}.

The magnetic chiral phase-diagram, $T$ vs. $eB$, for strange and light quarks calculated from the non-normalized approach given by Eq. (\ref{potential}) is given in the right-hand panel (b) of  Fig. \ref{fig:TP1}. It is clear that $T_c $ increases with the external magnetic field for both strange and light quark flavors. These results agree well with various studies using effective QCD-like models \cite{Marco2010,Marco2011,Marco2011-2,Fraga2013,Jens13,Skokov2011}, and also with the lattice QCD calculations \cite{lattice-delia}.

\subsection{Other approaches}

\begin{figure}[htb]
\centering{
\includegraphics[width=5.cm,angle=-90]{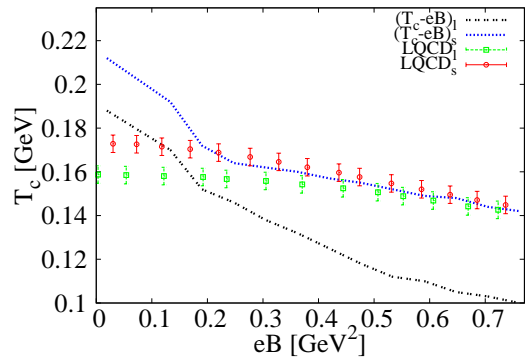}
\caption{(Color online)  The magnetic chiral phase-diagram, $T_c$ vs. $eB$, in an external magnetic field from the renormalized approach, Eq. (\ref{new-potential}), for strange (dotted curve), light quark chiral phase-transition (double dotted curve), compared with lattice QCD calculations \cite{Bali:2011qj} for light (open boxes) and strange quark (open circles) chiral phase-transition  \cite{Bali:2011qj} (dotted curve). \label{fig:TP2}}
}
\end{figure}

As mentioned before, in the system controlled by Eq. (\ref{potential}), there are some restrictions added to the quarks by  the magnetic field. This restrictions cause an increase in the transition temperature with increasing magnetic field. This means that  temperatures higher than $T_c$ should be needed to make the hadron-quark phase transition possible, where $T_c$ characterizes the critical temperature in absence of magnetic field. This interpretation or simply understanding seems to agree with most studies using QCD-like models, like PNJL and PLSM \cite{Marco2010,Marco2011,Marco2011-2,Fraga2013,Jens13,Skokov2011}. Furthermore, it fits well with the lattice QCD  calculations \cite{lattice-delia}. Accordingly, the transition temperature increases with the magnetic field. The renormalized approach, Eq. (\ref{new-potential}), leads to an inverse temperature evolution, i.e. $T_c$  decreases with increasing magnetic field. In Fig. \ref{fig:TP2}, we confront $T_c$ from the renormalized approach, Eq. (\ref{new-potential}) with the lattice QCD calculations \cite{Bali:2011qj}.

As shown in section \ref{sec:lqcdCom}, the main differences between the two sets of the lattice QCD calculations \cite{lattice-delia,Bali:2011qj} would be the quark flavors and masses besides lattice sizes and spacings. The dependence of $T_c$ on $eB/T^2$ in Ref. \cite{lattice-delia} is hard to be converted to the physical units we implement in the present work. The magnetic field $eB$ is related to some integer number through a factor function of the lattice spacing, $a$ and other variables such as, $eB=6\pi b T^2 (N_t/L_s)^2=6 \pi b (1/(a L_s))^2$, where $b$ is an integer, $N_t=1/(a\, T)$ is the lattice temporal dimension and $L_s$ is the spatial extents of the lattice measured in units of $a$.

Now we compare our results of  magnetic chiral phase-transition from the approach represented by Eq. (\ref{potential}) with $N_f=2+1$  flavors (two light plus one strange quarks) with the results from the SU(3) Polyakov NJL (PNJL) model \cite{Ferreira2013} and the entangled PNJL (EPNJL) \cite{Ferreira2013} at $N_f=2+1$ flavors (two light plus one strange quarks) with physical quarks and pion masses. The pseudocritical temperatures for $u$- and $d$-quark phase-transitions varies with increasing $eB$.

As introduced, the quark matter in strong magnetic fields can be studied in SU(3) PNJL model \cite{PNJL11,PNJL12,PNJL13}, in which SU(3) NJL model with scalar and pseudoscalar and t' Hooft six fermion interactions is included. The coupling between the magnetic field $B$ and the quarks and that between the effective gluon field and the quarks are implemented via the covariant derivative. To reproduce the lattice QCD results \cite{Ratti:2006}, an effective potential $\mathcal{U}\left(\phi,\bar\phi;T\right)$, where $\phi$ is the Polyakov-loop field in the algorithmic form is chosen. Also in the PNJL model, the coupling of scalar-type four-quark interaction in the NJL sector is taken into consideration. An effective vertex depending on the Polyakov loop $G(\phi, \bar{\phi})$ characterizes the EPNJL model.

\begin{figure}[htb]
\centering{
\includegraphics[width=5.cm,angle=-90]{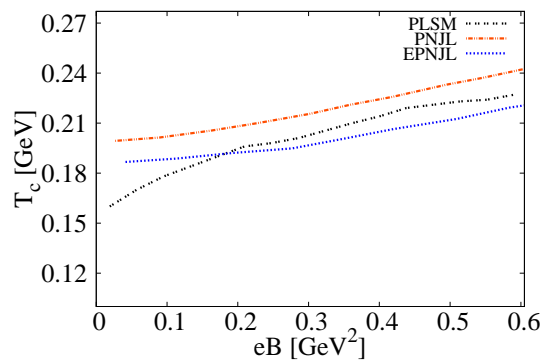}
\caption{(Color online)  The magnetic chiral phase-diagram, $T$ vs. $eB$, in an external magnetic field from the non-normalized approach, Eq(\ref{potential}), for light quark chiral phase-transition of PLSM (double dotted curve) , compared with PNJL (dash-dotted curve) and EPNJL (dotted curve) \cite{Ferreira2013}. \label{fig:TP3}}
}
\end{figure}

In Fig. \ref{fig:TP3}, the magnetic chiral phase-transition for light quarks in the  non-renormalized PLSM, Eq. (\ref{potential}), PNJL  and  EPNJL \cite{Ferreira2013} is represented by double-dotted, dash-dotted and dotted curve, respectively. In these three approaches, we find the same behavior, $T_c$ increases with increasing the external magnetic field.  Obviously, the increase in $T_c$ with increasing $eB$ as observed in both PNJL and EPNJL agrees with our result.

\section{Discussion and outlook}
\label{sec:disc}

In the present work,  we introduced two approaches for the magnetic effects on effective models based on the PLSM, Eqs. (\ref{potential}) and (\ref{new-potential}),. The effects of an external magnetic field are controlled by these approaches through the covariant derivative. The main difference between the two approaches is the Polyakov-loop field. In the approach represented by Eq. (\ref{potential}), the non-renormalized Polyakov loop and quark potential term are dominated. Through the magnetic field, there are some restrictions due to the Landau quantization added to the system, Eq. (\ref{potential}) or to the quarks energy. This enforces the saturation part of the pressure, for instance, to decrease with increasing $T$, right-hand panel (b) of Fig. \ref{fig:pr}.  In the approach represented by Eq. (\ref{new-potential}), a renormalized Polyakov-loop potential is inserted. Furthermore, we find that the value of the Polyakov-loop potential increases. Due to the magnetic field, there are restrictions added to the system, Eq.  (\ref{new-potential}) or to the quarks (energy), as well. This makes the saturated part of the pressure decreasing. Apparently, the subtraction is offset by the effect of the renormalized Polyakov loop, left-hand panel (a) of Fig. (\ref{fig:pr}).

In section \ref{sec:Results}, we introduced the thermal evaluation of the chiral condensates, the order parameters related the Polyakov-loop fields, some thermodynamic quantities, and the magnetic susceptibility under the effects of varying values of the external magnetic field. We described the chiral phase-transition in  dependence on the temperature. In doing this, we analysed the strange and non-strange chiral condensates and their first derivative  with respect to the temperature. This gives a signature for the magnetic chiral phase-transition in both systems represented by Eqs. (\ref{potential}) and (\ref{new-potential}). Also, we introduced the thermal evaluation of the Polyakov-loop field under the effect of external magnetic field. The thermal evaluation of the first derivative of the Polyakov loop is analysed. Then, we  introduced the magnetic properties including the susceptibility of the two systems. Finally we evaluated the magnetic chiral phase-transition.

The main conclusion from these results are summarized in Fig. \ref{fig:TP1}. The behavior of the magnetic chiral phase-transition for the system represented by Eq. (\ref{potential}),  the effect of the Landau quantization requires a reduction in the electromagnetic interactions, which makes the system requires more temperatures in order to move from the hadronic state to the QGP. This increases the critical temperature $T_c$, especially with increasing the external magnetic field. This dependence seems to agree well with the lattice QCD calculations \cite{lattice-delia} and with PNJL and EPNJL models \cite{Ferreira2013} as shown in Fig. (\ref{fig:TP3}). Also, in the system represented by renormalized approach, Eq. (\ref{new-potential}),  again the effect of the Landau quantization is assumed to reduce the electromagnetic interactions, this time due to the fact of increasing the color interactions and the dominant gluon potential.  In this case, the phase transition becomes fast with the increase magnetic filed. We find that  $T_c$ decreases with increasing magnetic field. This behavior agrees qualitatively well with the lattice QCD calculations \cite{Bali:2011qj} as shown in Fig. \ref{fig:TP2}.  

The ultimate goal of these studies is proposing quantities to be measured, experimentally, as signatures  for the effects of the external magnetic effect on the paramagnetic matter. In a future work, we plan to improve the present model, PLSM, in a way to get a much better agreement  with the lattice QCD calculations. 

The enhancement of the spontaneous breaking of the chiral symmetry in a non-Abelian gauge theory was predicted by the chiral perturbation theory for full QCD. In SU(2) Yang-Mills theory on lattice, it is found that the chiral condensate grows linearly with the field strength $B$ \cite{LatticeB1}. Also, the slope of the linear dependence should not be affected by the logarithmic volume dependence in the quenched limit \cite{LatticeB2}. But, increasing temperature decreases the coefficient in front of the linear term.  Recently \cite{Bali2014}, the consequences of the QCD paramagnetic properties were characterized as {\it chunks} of QGP that should become squeezed perpendicular to the magnetic field. This additional anisotropy should be subtracted from the measured elliptic flow $v_2=\langle\cos[2(\phi-\Phi)]\rangle$, where $\phi$ is the emission azimuthal angle with respect the reaction plane angle $\Psi$ in the  heavy-ion collisions. In doing this, we would unveil the flow due to the intrinsic fluid properties of the strongly interacting matter. According to \cite{Bali2014}, this effect is known as {\it ''paramagnetic squeezing''}. Furthermore, a pressure gradient due to the initial geometry should be taken into consideration. The temporal and spacial distribution of the external magnetic field \cite{reffA1,reffA2} and the early hydronization \cite{reffA3,reffA4} determines whether the geometric pressure gradient or paramagnetic squeezing becomes dominant.

The aim is to characterize the magnetic properties as a reflection to the way in which the system (matter) behaves under the effects of an external magnetic field. The magnetic properties is mainly represented by the magnetic susceptibility. As in solid state physics, the magnetic susceptibility, which is the second derivative for the free energy density of the system represented by Eq. (\ref{ch1}), depends strongly on essential magnetic properties. The magnetic susceptibility for both approaches given by Eqs. (\ref{potential}) and (\ref{new-potential}) is summarized in Fig. (\ref{fig:chim}). We find that the magnetic susceptibility is finite and positive. Furthermore, it  increases with the temperature. This reflects that both approaches, Eqs. (\ref{potential}) and (\ref{new-potential}), have almost the same magnetic property, i.e. paramagnetic. Also, we find the system represented by Eq. (\ref{new-potential}) is more close to the lattice QCD calculations \cite{Claudio2013} than the approach represented by Eq. (\ref{new-potential}). The agreement of the approach Eq. (\ref{new-potential}) with the lattice QCD calculations \cite{Claudio2013} means that the proposed approach is very close to the reality. The agreement of both approaches with the lattice QCD simulations \cite{Claudio2013} gives an important conclusion that both approaches represent a paramagnetic system but each system has its own mode of transmission from the hadronic state to the QGP state under the effects of an external magnetic field.

\section{Conclusions}
\label{sec:conclusion}

In peripheral heavy-ion collisions, an intrinsic magnetic field is likely produced. Its direction is upwards to the collision plane, duration is  very short ($\sim 1~$fm/c) but  magnitude is very strong ($\sim 10^{16}~$Tesla). In the present work, we concentrate the discussion on external magnetic field. The effects of the external magnetic field on the SU(3) PLSM for $N_f=2+1$ is studied. We introduced two approaches representing two types of Polyakov-loop fields. One of them is represented by Eq. (\ref{potential}), in which non-renormalized Polyakov-loop field in included. In this system, we find that the critical temperature $T_c$ increases with the external magnetic field. Apparently, this system has a paramagnetic property, i.e. attraction by the external magnetic field and positive magnetic susceptibility. This agrees well with most studies in the QCD-like effective models, like PNJL and SU(2) PLSM   \cite{Marco2010,Marco2011,Marco2011-2,Fraga2013,Jens13,Skokov2011}. The other system is represented by Eq. (\ref{new-potential}), in which the Polyakov loop field is renormalized. In this system, we find that the critical temperature $T_c$ significantly decreases with the increase in the magnetic field. Also, in this system we obtain a paramagnetic response. The magnitude increases with the temperature. The magnetic phase-transition of this system agrees well with the lattice QCD calculations \cite{Bali:2011qj}, especially for $s$-quark. But, the agreement becomes worth for the light quarks, especially at low values of the external magnetic field.


\appendix

\section{Mean Field  Approximation}
\label{appnd:1}

To calculate the grand potential in the mean field approximation,  we construct the partition function. In thermal equilibrium, the grand partition function can be defined by using a path integral over the quark, antiquark and meson field 
\begin{eqnarray} 
\mathcal{Z}&=& \mathrm{Tr \,exp}[-(\hat{\mathcal{H}}-\sum_{f=u, d, s}
\mu_f \hat{\mathcal{N}}_f)/T] \nonumber\\
&=& \int\prod_a \mathcal{D} \sigma_a \mathcal{D} \pi_a \int
\mathcal{D}\psi \mathcal{D} \bar{\psi} \mathrm{exp} \left[ \int_x
(\mathcal{L}+\sum_{f=u, d, s} \mu_f \bar{\psi}_f \gamma^0 \psi_f )
\right], \label{PF}
\end{eqnarray}
where $\int_x\equiv i \int^{1/T}_0 dt \int_V d^3x$, $V$ is the volume of the system and $\mu_f$ is the chemical potential for $f=u, d, s$. We take into consideration a symmetric quark matter and define a uniform blind chemical potential $\mu_f \equiv \mu_{u, d}=\mu_s$. The partition function is evaluated in the mean field  approximation \cite{Marco2010,Marco2011,Marco2011-2,Fraga2013,Jens13,Skokov2011,Schaefer:2007c,Schaefer:2008hk,blind}. Then, in the action, the meson fields are replaced by their expectation values $\bar{\sigma_x}$ and $\bar{\sigma_y}$ \cite{Kapusta:2006pm,Mao:2010}. We can use standard methods \cite{Kapusta:2006pm} to calculate the integration over the fermions yields. Then, by using the two Lagrangian in Eqs. (\ref{plsm}) and (\ref{plsm-new}) in the partition function, Eq. (\ref{PF}), we get two expressions for the thermodynamic potential.

\section{Magnetic catalysis}
\label{appnd:2}

For simplicity, we assume that the direction of the magnetic field goes along  $z$-direction. From the magnetic catalysis \cite{Shovkovy2013} and by using Landau quantization, we find that when the system is affected by a magnetic field. The quark dispersion relation  will be modified to be quantized by Landau quantum number, $n\geq 0$, and the concept of dimensional reduction will be applied
\begin{eqnarray}
E_u &=&\sqrt{p_{z}^{2}~+m_{q}^{2}~+|q_{u}|(2n+1-\sigma) B}, \label{Eu} \\
E_d &=&\sqrt{p_{z}^{2}~+m_{q}^{2}~+|q_{d}|(2n+1-\sigma) B}, \label{Ed} \\
E_s &=&\sqrt{p_{z}^{2}~+m_{s}^{2}~+|q_{s}|(2n+1-\sigma) B}, \label{Es}
\end{eqnarray} 
where $n$ is Landau quantum number and $\sigma $ is related to the spin quantum number, $S$ ($\sigma=\pm S/2$). Here,  we replace $2n+1-\sigma$ by one quantum number $\nu$, where $\nu=0$ is the Lowest Landau Level (LLL) and the Maximum  Landau Level (MLL) was determined according to Eq. (\ref{MLL}) \cite{Sutapa}, $ m_{f} $ where $f$ run over $u$, $d$ and $s$ quark masses, respectively, can be as, 
\begin{eqnarray}
m_q &=& g \frac{\sigma_x}{2}, \label{qmass} \\
m_s &=& g \frac{\sigma_y}{\sqrt{2}}.  \label{sqmass}
\end{eqnarray} 

We apply another magnetic catalysis property \cite{Shovkovy2013}, namely the dimensional reduction. As the name says, the dimensions will be reduced as $D\longrightarrow D-2$. In this situation, the three-momentum integral will transformed into a one-momentum integral
\begin{eqnarray} 
T~\int \dfrac{d^3 p}{(2 \pi)^3}~ \longrightarrow~  \dfrac{|q_{f}| B T}{2 \pi} \sum_{\nu=0}^{\infty} \int \dfrac{d p}{2 \pi} (2-1 \delta_{0n}). \label{DR}
\end{eqnarray} 
when $2-1 \delta_{0n}$ represents the degenerate in the Landau level, since for LLL we have single degenerate and doublet for the upper  Landau levels, 
\begin{eqnarray} \label{MLL}
\nu_{max} &=& \dfrac{\Lambda_{QCD}^{2}}{2 |q_{f}| B}.
\end{eqnarray} 
We use $m_q$ and $m_s$ for non-strange and strange quark mass, i.e. the masses of light quarks degenerate. This is not the case for the electric charges. In section \ref{sec:approaches}, $q_u$, $q_d$ and $q_s$ are elaborated.

\section{Minimization condition}
\label{appnd:3}

We notice that the thermodynamic potential density as given in Eqs. (\ref{potential}) and (\ref{new-potential}), has seven parameters $m^2, h_x, h_y, \lambda_1, \lambda_2, c$ and $g$ two unknown condensates $\sigma_x$ and $\sigma_y$ and an order parameter for the deconfinement, $\phi$ and $\phi^*$ or $\phi_R$ and $\phi^{*}_{R}$. The six parameters $m^2, h_x, h_y, \lambda_1, \lambda_2 $ and $ c$  are fixed in the vacuum by six experimentally known quantities \cite{Schaefer:2008hk}. In order to evaluate the unknown parameters $\sigma_x$, $ \sigma_y$, $\phi$ and $\phi^*$ or $\phi_R$ and $\phi^{*}_{R}$, we can minimize the thermodynamic potential, Eqs. (\ref{potential}, \ref{new-potential}), with respect to $\sigma_x$, $\sigma_y$, $\phi$ and $\phi^*$ or $\phi_R$ and $\phi^{*}_{R}$. Doing this, we obtain a set of four equations of motion
\begin{eqnarray}
\left.\frac{\partial \Omega_{1}}{\partial \sigma_x}= \frac{\partial
\Omega_{1}}{\partial \sigma_y}= \frac{\partial \Omega_{1}}{\partial
\phi}= \frac{\partial \Omega_{1}}{\partial \phi^*}\right|_{min} =0, \label{cond1}\\
\left.\frac{\partial \Omega_{1}}{\partial \sigma_x}= \frac{\partial
\Omega_{1}}{\partial \sigma_y}= \frac{\partial \Omega_{1}}{\partial
\phi_{R}}= \frac{\partial \Omega_{1}}{\partial \phi^{*}_{R}}\right|_{min} =0, \label{cond2}
\end{eqnarray}
meaning that $\sigma_x=\bar{\sigma_x}$, $\sigma_y=\bar{\sigma_y}$, $\phi=\bar{\phi}$ and $\phi^*=\bar{\phi^*}$ or $\phi_R$ and $\phi^{*}_{R}$ are the global minimum.



\end{document}